\documentclass[%
 reprint,
 superscriptaddress,
 amsmath,amssymb,
 aps,
 prb,
 longbibliography,
]{revtex4-2}

\usepackage{graphicx}
\usepackage{dcolumn}
\usepackage{bm}
\usepackage{physics}
\usepackage[colorlinks=true,citecolor=blue,linkcolor=blue,urlcolor=blue]{hyperref}
\usepackage{mathtools}
\usepackage[dvipsnames]{xcolor}
\usepackage[normalem]{ulem} 
\usepackage{cancel}

\newcommand{\omegad}[0]{\omega_\textrm{d}}
\newcommand{\omegaL}[0]{\omega_\textrm{L}}
\newcommand{\OmegaR}[0]{\Omega_\textrm{R}}
\newcommand{\Hso}[0]{H_\textrm{so}}
\newcommand{\lso}[0]{\ell_\textrm{so}}
\newcommand{\ld}[0]{\ell_\textrm{d}}

\begin{document}

\title{Rabi oscillations induced by nonresonant gyrating drive}

\author{Baksa Kolok}
\affiliation{%
 Department of Theoretical Physics, Institute of Physics, Budapest University of Technology and Economics, M\H{u}egyetem rkp. 3., H-1111 Budapest, Hungary
}%
\affiliation{HUN-REN-BME-BCE Quantum Technology Research Group, Budapest University of Technology and Economics, M\H{u}egyetem rkp. 3., H-1111 Budapest, Hungary}

\author{Andr\'as P\'alyi}
\affiliation{%
 Department of Theoretical Physics, Institute of Physics, Budapest University of Technology and Economics, M\H{u}egyetem rkp. 3., H-1111 Budapest, Hungary
}%
\affiliation{HUN-REN-BME-BCE Quantum Technology Research Group, Budapest University of Technology and Economics, M\H{u}egyetem rkp. 3., H-1111 Budapest, Hungary}

\date{\today}

\begin{abstract}
Coherent control of two-level quantum systems is typically achieved using resonant driving fields, forming the basis of qubit operations.
Here, we report a mechanism for inducing complete Rabi oscillations in monochromatically driven two-level systems even when the drive frequency is much larger than the qubit’s Larmor frequency.
This effect---Rabi oscillations induced by nonresonant gyrating drive (RING)---requires a gyrating control field, i.e., an elliptic drive trajectory that encloses a nonzero area per cycle.
We illustrate the effect with numerical simulations and provide an analytical understanding via a simple effective Hamiltonian derived from Floquet theory and perturbation theory.
We show that RING enables coherent oscillations without relying on resonant energy exchange, thereby allowing the integration of high-pass noise filtering.
We detail a realization in electrically driven spin-orbit qubits, where the mechanism provides access to non-Abelian phases at nonzero magnetic field and enables amplification of the Rabi frequency using the same gate voltage amplitudes at higher drive frequencies.
Our results broaden the landscape of quantum control by establishing a route to coherent manipulation under strongly off-resonant driving conditions.
\end{abstract}

\maketitle

\section{Introduction}

Coherent control of two-level quantum systems is typically achieved via Rabi oscillations induced by resonant driving \cite{rabi1937rabioscillation}.
This mechanism enables high-fidelity qubit operations in various platforms, including superconducting circuits \cite{kjaergaard2020superconducting, Nakamura1999CoherentBox, Vion2002ManipulatingCircuit, chiorescu2003coherent, wallraff2004strong, manucharyan2009fluxonium}, trapped ions \cite{harty2014high, ballance2016high}, semiconductor spins \cite{Burkard2023SemiconductorQubits, Koppens2006DrivenDot, Nowack2007CoherentFields, Pla2012ASilicon, Veldhorst2014AnControl-fidelity, Crippa2018ElectricalQubits, Froning2021UltrafastFunctionality, Hendrickx2021AProcessor, Philips2022UniversalSilicon, Wang2024PursuingDot, Pla2013High-fidelitySilicon, Tettamanzi2017ProbingFrequencies, Hile2018AddressableSilicon, Koch2019SpinTransistor, alegre2007polarization, london2014strong}, and cold atoms \cite{Schrader2004NeutralAtoms, Wang2015NeutralAtomIndividual, Xia2015NeutralAtomsRandBench}.

While resonant Rabi oscillations provide robust control, other strategies have emerged to address specific challenges.
For instance, in singlet-triplet qubits \cite{Levy2002Singlettriplettheory, Petta2005singlettripletfirst, Maune2012singlettriplet, Wu2014twoaxiscontrol, Jirovec2021singlettripletGe, Zhang2025foursinglettriplet} and exchange-only qubits \cite{Divincenzo2000exchangeonlytheory1, Kempe2001exchangeonlytheory2, Weinstein2023exchangeonlyexp, Bosco2024exchangeonlyspinorbitqubitssilicon}, operations are driven by baseband pulses that modify the exchange interactions between spins.
Another example is 
baseband control of hopping single-spin qubits in semiconductor quantum dots \cite{unseld2024basebandcontrolsingleelectronsilicon, wang2024operating, rimbachruss2025gaplessspinqubit}, which alleviates the need for microwave control fields.
A third example is holonomic quantum control, a scheme that provides partial tolerance against control errors \cite{falci2000detection,AbdumalikovJr2013, leibfried2003experimental, toyoda2013realization, arroyo2014room, zu2014experimental, yale2016optical, sekiguchi2017optical, ishida2018universal, zhou2017holonomic, nagata2018universal, hong2018implementing}, and, in case of semiconductor quantum dot spin qubits, alleviates the need for a magnetic field \cite{golovach2010holonomicspinqubits, San-Jose2008GeometricDecoherence, kolok2024protocols}.

Here, we propose and theoretically demonstrate a control mechanism that enables complete Rabi oscillations without resonance; more precisely, by utilizing a monochromatic drive signal with a frequency exceeding the Larmor frequency. 
This effect, which we call Rabi oscillations induced by nonresonant gyrating drive (RING), emerges when the qubit is driven by a gyrating field, i.e., a circular or elliptic drive, which traces a finite-area loop in control space. 
The resulting oscillations broaden the quantum control toolbox by exploiting off-resonant dynamics.

The RING mechanism introduced here exhibits several distinctive features that set it apart from conventional qubit control schemes. It provides a route to demonstrating non-Abelian phases, in the context of spin-orbit coupled semiconductor spin qubits, in the presence of a nonzero magnetic field, which is of fundamental interest and relevant for the development of holonomic quantum computation \cite{San-Jose2008GeometricDecoherence, golovach2010holonomicspinqubits, kolok2024protocols}.
However, we emphasize that while the RING mechanism shares the presence of closed trajectories in control space with adiabatic and nonadiabatic geometric approaches, the physical mechanism and operational regime are fundamentally different.
Conventional adiabatic geometric control  protocols require degeneracy \cite{zanardi1999holonomic, golovach2010holonomicspinqubits}. 
Nonadiabatic geometric approaches
typically rely on engineered multilevel structures, auxiliary states, and controlled phase accumulation, while canceling out the dynamical phase \cite{sjoqvist2012nonadiabaticholonomic, AbdumalikovJr2013, zu2014experimental}.
In contrast, the RING mechanism operates in a simple monochromatically driven two-level system and does not rely on adiabaticity, degeneracy, or auxiliary states.

Moreover, by achieving full Rabi oscillations with strongly detuned driving, this scheme naturally permits the use of high-pass filtering to attenuate low-frequency and on-resonance noise. An analogous strategy was implemented using subharmonic driving, which instead supports the incorporation of low-pass filtering for fluxonium \cite{schirk2025subharmonicfluxonium}. Finally, the mechanism operates without relying on resonant energy exchange between the drive and the qubit, in contrast to standard resonant control protocols.
At the same time, the RING drive can induce a Rabi frequency that is comparable to the Larmor frequency of the qubit, which opens the door to fast qubit manipulation for low-frequency qubits. 

The rest of this paper is structured as follows. In Sec.~\ref{sec:ring} we present a minimal two-level model with elliptic drive, and demonstrate the RING mechanism numerically and analytically using Floquet and quasidegenerate perturbation theory. 
In Sec.~\ref{sec:ringed} we show that the non-Abelian Berry phase makes the RING drive possible in spin-orbit coupled semiconductor quantum dot devices.
Section~\ref{sec:discussion} surveys experimental platforms and implementation challenges (pulse shaping, detuning robustness, and error mitigation). 
Finally, Sec.~\ref{sec:conclusion} summarizes our findings and outlines prospects for near-term demonstrations.

\section{RING Mechanism in a Driven Two-Level System}
\label{sec:ring}

In this section, we provide a minimal model for the RING mechanism.
We consider a two-level system with a nonzero energy splitting (Larmor frequency $\omega_\mathrm{L}/2\pi$), which is driven by two linearly polarized control fields that are orthogonal to each other, have the same driving frequency $\omega_\mathrm{d}/2\pi$, and have a phase lag with respect to each other. 
This drive produces complete Rabi oscillations even if the drive frequency is much higher than the Larmor frequency.

To illustrate the effect, we first present numerical results for a rotating drive field.
Then, we explain the observed dynamics using Floquet theory.

\subsection{Numerical demonstration of the RING mechanism}

\begin{figure}[t]
\centering
\includegraphics[width=\linewidth]{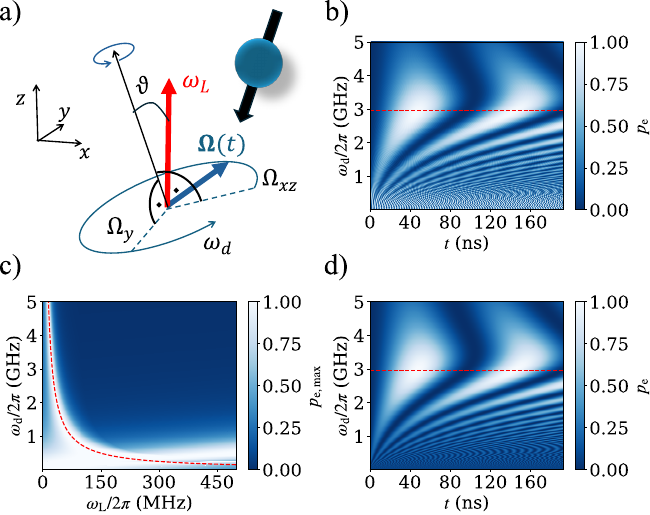}
\caption{Rabi oscillation induced by nonresonant gyrating drive in a two-level system. 
(a) Schematics of the driving scheme inducing RING in a two-level system, represented by a spin (sphere with arrow). 
The static component of the Zeeman field ($\omegaL$, red arrow) encloses a nonzero angle $\vartheta$ with the normal of the plane spanned by the drive fields ($\boldsymbol{\Omega}(t)$, blue arrow).
(b) Excited-state population $p_\mathrm{e}$ as a function of time and drive frequency.
Initial state is the ground state $\ket{\downarrow}$ of the static part of the Hamiltonian \eqref{eq:Hamiltonian}, and the system is driven as specified by Eq.~\eqref{eq:Hamiltonian}. 
Red dashed line: drive frequency corresponding to full Rabi oscillation.
The data are obtained by numerically solving the Schrödinger equation; see the Appendix. 
Drive strengths: $\Omega_{xz}/2\pi = \Omega_y/2\pi = 400$~MHz.
Tilt angle: $\vartheta = \pi/8$.
Larmor frequency: $\omegaL/2\pi = 25$~MHz. 
Phases: $\phi_{xz} = 0$ and $\phi_y = -\pi/2$.
(c) Maximum excited-state population as a function of Larmor and drive frequencies, with the same parameters as in (b). 
For each point, $p_\mathrm{e}(t)$ was maximized in a time window of duration $T = 1000 \times 2\pi/\omegad$.
Besides the fundamental resonance at $\omegad = \omegaL$ (and the fainter half-harmonic resonance at $\omegad = \omegaL/2$), full population inversion occurs in the $\omegad \gg \omegaL$ regime, with $\sim 1/\omegaL$ behavior.
We have plotted the ``effective resonance'' condition of Eq.~\eqref{eq:ring_freq} as a red dashed line.
(d)  Excited-state population as a function of time and drive frequency, calculated analytically as in Eq.~\eqref{eq:Rabioscillation}, with the same parameter values as (c).
We marked the frequency of full Rabi oscillations with red dashed line, using Eq.~\eqref{eq:ring_freq}.}
\label{fig:ring}
\end{figure}

The Hamiltonian of the two-level system with elliptically polarized drive reads as
\begin{multline}
H(t) =\frac{\hbar \omegaL}{2} \sigma_z + \frac{\hbar\Omega_{xz}}{2}  \cos(\omegad t + \phi_{xz}) (\cos\vartheta \sigma_x  + \sin\vartheta \sigma_z) \\ + \frac{\hbar\Omega_y}{2} \cos(\omegad t + \phi_y) \sigma_y,
\label{eq:Hamiltonian}
\end{multline}
where \(\Omega_{xz}, \Omega_y\) are driving amplitudes, \(\phi_{xz}, \phi_y\) are initial phases of the drive components, and \(\vartheta\) is the angle between the static field and the normal vector of the driving field's plane, see Fig.~\ref{fig:ring}(a).

Figures~\ref{fig:ring}(b) and \ref{fig:ring}(c) show the key result: complete Rabi oscillations, despite the drive frequency \(\omegad\) being far detuned from the Larmor frequency \(\omegaL\): In particular, Fig.~\ref{fig:ring}(b) exhibits Rabi oscillations around $\omegad/2\pi \approx 3\, \mathrm{GHz}$, while $\omegaL / 2\pi = 25$ MHz. 
In this numerical demonstration, we set $\Omega_{xz} = \Omega_y$ and $\phi_{xz} = 0, 
\phi_y = -\pi/2$ for simplicity, and we set the initial state as the ground state of the static part of the Hamiltonian.
(See caption of Fig.~\ref{fig:ring} for further parameter values.)
In Fig.~\ref{fig:ring}(b), the color scale shows the excited-state occupation probability $p_\mathrm{e}$ as the function of time and drive frequency. 
The excited state probability shows a chevronlike pattern similar to the case of a resonant drive;
however, there are clear differences:
(i) The pattern in Fig.~\ref{fig:ring}(b) has no mirror symmetry axis at any value of the drive frequency, and 
(ii) the drive frequency corresponding to complete Rabi oscillations, marked by red dashed line, is not the same as the drive frequency corresponding to the lowest oscillation frequency.

To obtain Fig.~\ref{fig:ring}(c), we simulated the time-dependent excited-state probability $p_\mathrm{e}(t)$ for various values of the Larmor and drive frequencies, and plotted the maximum value of $p_\mathrm{e}(t)$ as a function of the two frequencies. 
We observe the fingerprint of resonant Rabi oscillations in Fig.~\ref{fig:ring}(c) as the white band at the bottom of the graph following the linear trend $\omegad = \omegaL$.
Besides that standard feature, another ``effective resonance'' (full population inversion) occurs for driving frequencies highly above the Larmor frequency, following an $\omegad \propto 1/\omegaL$ condition, highlighted as the dashed red line.
This is the fingerprint of the RING mechanism.

\subsection{Analytical, perturbative description of the RING mechanism}

We provide an analytical, perturbative description of the phenomena identified in the numerical simulations. 
By combining the Floquet formalism \cite{shirley1965floquet, romhanyi2015subharmonic} with quasidegenerate perturbation theory \cite{bravyi2011schrieffer, bukov2015universal, winkler2003spinorbit}, we explicitly derive an effective Hamiltonian in the high-frequency limit.
While the dynamics of periodically driven systems in this regime have been studied extensively in the literature \cite{saar2003effectivehamiltonian, takahiro2016BWtheory}, the specific configuration considered here has not been analyzed. 
We therefore present the full derivation to make the underlying mechanism explicit and enable a transparent interpretation of the results.

For a periodically driven two-level system, Floquet’s theorem yields two solutions \cite{shirley1965floquet, romhanyi2015subharmonic} of the form $\ket{\Psi_\alpha(t)} = e^{-i\epsilon_\alpha t/\hbar}\ket{\Phi_\alpha(t)}$, where $\ket{\Phi_\alpha(t)}$ is periodic with the drive period and $\alpha = \pm$ labels the Floquet states corresponding to higher and lower quasienergy. 
We expand $\ket{\Phi_\alpha(t)}$ in Fourier harmonics:
\begin{equation}
\label{eq:floquetansatz}
    \ket{\Psi_\alpha(t)} = e^{-i\epsilon_\alpha t/\hbar}\sum_{\sigma\in\{\uparrow,\downarrow\}}\sum_{n=-\infty}^\infty e^{in\omegad t}c_{\alpha, \sigma n}\ket{\sigma},
\end{equation}
where $\ket{\sigma}$ denotes the eigenbasis of the undriven Hamiltonian, i.e., $\sigma_z$. 

Substituting the trial solution Eq.~\eqref{eq:floquetansatz} to the time-dependent Schrödinger equation, one obtains an infinite-dimensional eigenvalue problem 
\begin{equation} \label{eq:floquetequation}
\bm{\mathcal{F}} \mathbf{c}_\alpha = \epsilon_\alpha \mathbf{c}_\alpha    
\end{equation}
for the Floquet matrix $\bm{\mathcal{F}}$. 
The notation $\mathbf{c}_\alpha$ stands for an infinite vector of the coefficients in Eq.~\eqref{eq:floquetansatz}, ordered as $\mathbf{c}_\alpha = (\dots, 
c_{\alpha,\uparrow,-1},
c_{\alpha,\downarrow,-1},
c_{\alpha,\uparrow,0}, \dots
)^T$.
For the Hamiltonian in Eq.~\eqref{eq:Hamiltonian}, the Floquet matrix reads as
\begin{equation}\label{eq:floquetmatrix}
    \boldsymbol{\mathcal{F}} = \begin{pmatrix}
        \ddots & \vdots & \vdots & \vdots &  \\
        \cdots &  H^{(0)} - \hbar \omegad & H^{(-1)} & 0 & \cdots \\
        \cdots & H^{(1)} & H^{(0)} & H^{(-1)} & \cdots\\
        \cdots&0 & H^{(1)} & H^{(0)} + \hbar \omegad & \cdots \\
         & \vdots & \vdots & \vdots & \ddots
    \end{pmatrix},
\end{equation}
where each nonzero entry is a $2\times 2$ block corresponding to the nonzero Fourier components of the Hamiltonian:
\begin{subequations}
\begin{align}
    H^{(0)} &= \frac{\hbar\omegaL}{2}\sigma_z,\\
    H^{(\pm 1)} &= \frac{\hbar}{4}\left(e^{\pm i\phi_{xz}}\Omega_{xz}(\cos\vartheta\sigma_x + \sin\vartheta\sigma_z) + e^{\pm i\phi_y}\Omega_y\sigma_y\right).
\end{align}
\end{subequations}

The eigenvalue equation for the Floquet matrix in Eq.~\eqref{eq:floquetequation} has infinitely many solutions, however, only two of the solutions are nonequivalent, e.g., those with quasienergy $\epsilon_\alpha$ lying in the first ``quasienergy Brillouin zone'': $-\hbar\omegad/2 < \epsilon_\alpha \le \hbar\omegad/2$ \cite{shirley1965floquet, romhanyi2015subharmonic}. 

\begin{figure}
    \centering
    \includegraphics[width=\linewidth]{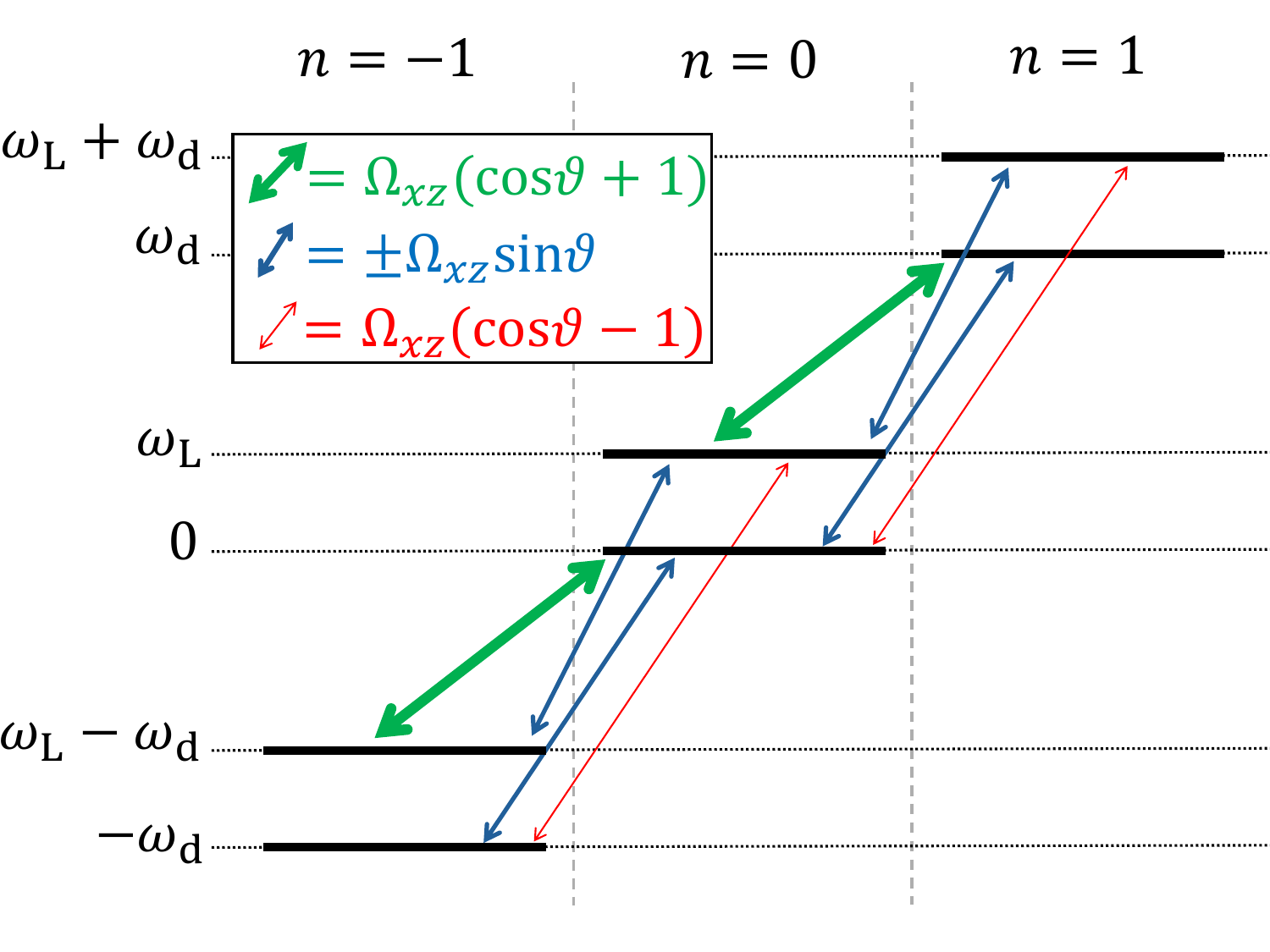}
    \caption{Structure of the Floquet matrix. 
    The figure shows three adjacent Fourier sectors ($n = -1, 0, 1$) of the Floquet ladder, each containing two unperturbed eigenstates split by the Larmor frequency $\omegaL$. 
    Arrows represent drive-induced couplings between states, when the drive is circular (\(\Omega_{xz} = \Omega_y, \phi_{xz}-\phi_{y} = \pi/2\)); their thickness reflects the magnitude of the matrix elements, which depends on the angle $\vartheta$. 
    For $0 < \vartheta < \pi/2$, this thickness hierarchy is preserved, resulting in a shift of energy levels that reduces the effective level splitting. 
    For $\vartheta > \pi/2$, the hierarchy reverses.
    Within $0 < \vartheta < \pi/2$, this structure allows for an ``effective resonance'', defined in Eq.~\eqref{eq:effresonance}, in the $(\Omega_{xz}=\Omega_y, \omegad)$ parameter space where Rabi oscillations are complete.
    Blue lines do not shift energy levels but enable coherent transitions; without them, Rabi oscillations vanish.}
    \label{fig:floquet}
\end{figure}

The structure of the Floquet matrix is illustrated in Fig.~\ref{fig:floquet}, in the special case when $\Omega_{xz} = \Omega_y$ and $\phi_{xz} - \phi_y = \pi/2$, i.e.,~for circular drive. 
Each horizontal line represents an ``unperturbed'' eigenvalue-eigenvector pair (or \emph{Floquet level}) of the system, that is, an eigenvalue-eigenvector pair of the diagonal part of $
\boldsymbol{\mathcal{F}}$.
These Floquet levels are
organized in Fig.~\ref{fig:floquet} into Fourier sectors labeled by $n = -1, 0, 1$. 
The harmonic drive introduces couplings between Floquet levels in neighboring sectors, shown as arrows in Fig.~\ref{fig:floquet}.
The thickness of each arrow indicates the magnitude of the corresponding matrix element; for this figure, we assume $0 < \vartheta < \pi/2$.
(Note that for $\vartheta > \pi/2$, while keeping $\phi_{xz} - \phi_y \in [0,\pi]$, the ordering of the green and red matrix-element magnitudes in Fig.~\ref{fig:floquet} reverses, which implies that no Rabi oscillation will occur.)

In the high-frequency drive limit, where \(\omegad \gg \omegaL,\Omega_{xz},\Omega_{y}\), we use quasidegenerate perturbation theory \cite{bravyi2011schrieffer, bukov2015universal, winkler2003spinorbit} to obtain an effective $2\times 2$ Floquet matrix from $\boldsymbol{\mathcal{F}}$, enabling to express the perturbed eigenvectors of $\boldsymbol{\mathcal{F}}$. 
This approach yields the following effective Floquet matrix, up to second order in 
$\omegaL$, $\Omega_{xz}$, and $\Omega_y$:
\begin{equation} \label{eq:Feff}
     \mathcal{F}_{\text{eff}} = H^{(0)} + \frac{[H^{(1)},H^{(-1)}]}{\hbar\omegad} = \frac{\hbar}{2}\left(\Delta\sigma_z + \OmegaR\sigma_x\right).
\end{equation}
Here, we introduced
\begin{subequations}
\label{eq:effectiveparameters}
\begin{align} \label{eq:effsplitting}
    \Delta &= \omegaL - \frac{\Omega_{xz}\Omega_y\cos\vartheta\sin(\Delta\phi)}{2\omegad}, \\
    \OmegaR &= \frac{\Omega_{xz}\Omega_y\sin\vartheta\sin\left(\Delta\phi\right)}{2\omegad}, \label{eq:effectivecoupling}
\end{align}
\end{subequations}
and $\Delta \phi = \phi_{xz}-\phi_y$.

The structure of the effective Floquet matrix can be interpreted in terms of second-order energy level shifts of the states $\ket{0\uparrow}$ and $\ket{0\downarrow}$ and the induced coupling between them. 
These level shifts and the coupling mechanism can be understood using the graphical representation of the Floquet matrix shown in Fig.~\ref{fig:floquet}. 

The $\ket{0\uparrow}$ state is coupled to four states: $\ket{1\downarrow}$ (green arrow), $\ket{-1\downarrow}$ (red arrow), and to the $\ket{\pm1 \uparrow}$ states (blue arrows).
All of these couplings result in a second-order energy shift due to level repulsion.
While the couplings to the $\ket{\pm1 \uparrow}$ states cancel each other, the level repulsion between $\ket{0\uparrow}$ and $\ket{1\downarrow}$ (green arrow) is dominant compared with the repulsion between $\ket{0\uparrow}$ and $\ket{-1\downarrow}$ (red arrow) in the parameter regime $0 < \vartheta < \pi/2$.
As a consequence, the net second-order energy correction to the $\ket{0\uparrow}$ state is negative. 
By an analogous reasoning, the energy of the $\ket{0\downarrow}$ state acquires a net positive shift. Hence, the energy gap between these two states is effectively reduced, in agreement with Eq.~\eqref{eq:effsplitting}.
Simultaneously, the combination of transverse and longitudinal couplings (green/red arrow and blue arrow) generates a second-order coupling between the states $\ket{0 \downarrow}$ and $\ket{0 \uparrow}$, which yields the parameter $\OmegaR$ in Eq.~\eqref{eq:effectivecoupling}.

The time-evolved Floquet states $\ket{\Psi_\alpha(t)}$ can be expressed via the diagonalization of the effective Floquet matrix $\mathcal{F}_{\text{eff}}$ of Eq.~\eqref{eq:Feff}, and the unitary transformation corresponding to the Schrieffer--Wolff transformation $e^S$, which is used to perturbatively eliminate the off-diagonal blocks between the zeroth and the other Fourier sectors. 

One obtains the leading order dynamics in the small parameters $\omega_\mathrm{L}/\omega_\mathrm{d}$,
$\Omega_{xz}/\omega_\mathrm{d}$,
and 
$\Omega_{y}/\omega_\mathrm{d}$ when the Schrieffer--Wolff unitary $e^S$ is approximated with the identity. 
In this approximation, $c_{\alpha,\sigma n}$ is only nonzero if $n=0$, and the two nonzero coefficients form the eigenvectors of the effective Floquet matrix $\mathcal{F}_{\text{eff}}$, i.e.,
\begin{eqnarray}\label{eq:c0}
    c_{\alpha,0\sigma} = f_{\alpha\sigma},
\end{eqnarray} 
such that the vectors
\begin{eqnarray}\label{eq:Feigen}
    \ket{f_\alpha} = \sum_{\sigma \in\{\uparrow,\downarrow\}} f_{\alpha\sigma} \ket{\sigma}  
\end{eqnarray}
are the eigenstates of $\mathcal{F}_{\text{eff}}$ for $\alpha = \pm$. 
The corresponding eigenvalues are the quasienergies $\epsilon_\alpha$.
Thus, the Floquet states, using Eq.~\eqref{eq:floquetansatz}, read as
\begin{eqnarray}\label{eq:floquetsol0}
    \ket{\Psi^{(0)}_\alpha(t)} &=& e^{-\frac{i}{\hbar}\epsilon_\alpha t}\ket{f_\alpha} =e^{-\frac{i}{\hbar} \mathcal{F}_\textrm{eff} t}\ket{f_\alpha}.
\end{eqnarray}

The time-evolution operator reads as
\begin{eqnarray} \label{eq:Ut0}
    U^{(0)}(t) = \sum_{\alpha \in\{+,-\}}\ketbra{\Psi^{(0)}_\alpha(t)}{\Psi^{(0)}_\alpha(0)} = e^{-\frac{i}{\hbar} \mathcal{F}_\textrm{eff} t},
\end{eqnarray}
where we used Eq.~\eqref{eq:floquetsol0} and the fact that eigenvectors of the effective Floquet matrix $\ket{f_\alpha}$ form a basis.
Finally, the excited-state population, when the system is initialized in $\ket{\downarrow}$, reads as
\begin{multline} \label{eq:Rabioscillation}
    p_\textrm{e}(t) = 
    |\bra{\uparrow}e^{-\frac{i}{\hbar} \mathcal{F}_\textrm{eff} t}\ket{\downarrow}|^2   
    \\
    =\frac{\OmegaR^2}{\OmegaR^2 + \Delta^2} \sin^2\left(\frac{\sqrt{\OmegaR^2 + \Delta^2}}{2}t\right),
\end{multline}
with the parameters defined in Eqs.~\eqref{eq:effectiveparameters}.

The ``effective resonance'' condition, when the system performs complete Rabi oscillations, is fulfilled when $\Delta=0$, i.e., from Eq.~\eqref{eq:effsplitting} it follows that 
\begin{eqnarray} \label{eq:effresonance}
    \omegaL &=& \frac{\Omega_{xz}\Omega_y\cos\vartheta\sin\left(\Delta\phi\right)}{2\omegad}.
\end{eqnarray}
This should be distinguished from a conventional resonance condition: The applied drive is far detuned from the bare Larmor transition, \(\omegad\gg\omegaL\), and Eq.~\eqref{eq:effresonance} does not describe a crossing between different bare Floquet sectors. Rather, the gyrating drive induces second-order Floquet level shifts that renormalize the splitting within the effective \(n=0\) Floquet sector. When this renormalized splitting vanishes, the off-diagonal term \(\OmegaR\) hybridizes the two effective states and enables complete Rabi oscillations.
Note that at the effective resonance, it holds that $\omegaL \ll \Omega_{xz},\Omega_{y}$. 
We call the drive frequency satisfying the effective resonance condition in Eq.~\eqref{eq:effresonance} as the RING frequency and denote it as 
\begin{eqnarray} \label{eq:ring_freq}
    \omega_\textrm{RING} = \frac{\Omega_{xz}\Omega_y\cos\vartheta\sin\left(\Delta\phi\right)}{2\omegaL}.
\end{eqnarray}

Another interesting consequence of Eq.~\eqref{eq:effresonance} is that the Rabi frequency is proportional to the Larmor frequency and not to the driving field amplitude when the effective resonance condition is fulfilled:
\begin{eqnarray} \label{eq:fullRabifreq}
    \OmegaR &=& \omegaL \tan\vartheta,
\end{eqnarray}
where we used Eqs.~\eqref{eq:effectivecoupling} and \eqref{eq:effresonance}.
Practically this means that, unlike in the resonant drive case, by increasing the drive amplitude one cannot increase the Rabi frequency, it just detunes the RING frequency, see Eq.~\eqref{eq:ring_freq}. The Rabi frequency, for a given Larmor frequency and angle $\vartheta$, is fixed as long as the effective resonance condition in Eq.~\eqref{eq:effresonance} is fulfilled.

In Fig.~\ref{fig:ring}(d), we have plotted the induced dynamics, in terms of the excited-state probability $p_\mathrm{e}$, using Eqs.~\eqref{eq:Rabioscillation} and \eqref{eq:effectiveparameters}.
This approximate analytical solution in Fig.~\ref{fig:ring}(d) agrees with the numerical results in Fig.~\ref{fig:ring}(b) up to fine structure. 

The fine structure of the numerical dynamics can be estimated from the first-order approximation of the unitary transformation corresponding to the Schrieffer--Wolff transformation: $e^{S} = I + S$. The $S$ matrix, up to first order, reads as
\begin{equation}
    S =  \begin{pmatrix}
        \ddots & \vdots & \vdots & \vdots &  \\
        \cdots & 0 & \frac{H^{(-1)}}{\hbar\omegad} & 0 & \cdots \\
        \cdots & -\frac{H^{(1)}}{\hbar\omegad} & 0 & \frac{H^{(-1)}}{\hbar\omegad} & \cdots \\
        \cdots & 0 & -\frac{H^{(1)}}{\hbar\omegad} & 0 & \cdots \\
         & \vdots & \vdots & \vdots & \ddots
    \end{pmatrix},
\end{equation}
where all the blocks indicated by dots are zeros. One obtains the $\mathbf{c}_\alpha$ vector by acting with $I + S$ on the zeroth order vector which has nonzero elements in the zeroth Fourier sector. The resulting vector has six nonzero elements in the $n=0$ and $n=\pm 1$ Fourier sectors. The $n=0$ coefficients remain the same as in Eq.~\eqref{eq:c0}. The other two two-element blocks read as
\begin{eqnarray}\label{eq:fourierblocks}
    \begin{pmatrix}
        c_{\alpha \uparrow, \pm1}\\
        c_{\alpha \downarrow, \pm1}
    \end{pmatrix} = \mp\frac{H^{(\pm1)}}{\hbar\omegad} \begin{pmatrix}
        f_{\alpha \uparrow}\\
        f_{\alpha \downarrow}
    \end{pmatrix},
\end{eqnarray}
where $f_{\alpha\uparrow}$ ($f_{\alpha\downarrow}$) is the first (second) component of the eigenstate of the effective Floquet matrix, see Eq.~\eqref{eq:Feigen}.
Therefore, the Floquet states, reconstructed
from Eq.~\eqref{eq:floquetansatz} using Eqs.~\eqref{eq:floquetsol0} and \eqref{eq:fourierblocks}, read as
\begin{equation} \label{eq:floquetsol}
    \ket{\Psi_{\alpha}(t)} =  \left(I + \tilde{S}(t)\right)\ket{\Psi^{(0)}_\alpha(t)},
\end{equation}
where 
\begin{equation}
    \tilde{S}(t) = \frac{e^{-i\omegad t} H^{(-1)} - e^{i\omegad t} H^{(1)}}{\hbar \omegad}.    
\end{equation}

The time-evolution operator, expressed with the Floquet states and using Eqs.~\eqref{eq:Ut0} and \eqref{eq:floquetsol} and neglecting second-order terms, reads as
\begin{multline} \label{eq:timeevolution}
    U(t) = \sum_{\alpha \in\{+,-\}} \ket{\Psi_\alpha(t)}\bra{\Psi_\alpha (0)} \\
    = e^{-\frac{i}{\hbar} \mathcal{F}_\textrm{eff} t} + \tilde{S}(t)e^{-\frac{i}{\hbar} \mathcal{F}_\textrm{eff} t} - e^{-\frac{i}{\hbar} \mathcal{F}_\textrm{eff} t}\tilde{S}(0).
\end{multline}

We express the correction to the excited-state population at the effective resonance ($\Delta = 0$) using Eq.~\eqref{eq:timeevolution} up to first order in the small parameters $\omega_\mathrm{L}/\omega_\mathrm{d}$,
$\Omega_{xz}/\omega_\mathrm{d}$,
and 
$\Omega_{y}/\omega_\mathrm{d}$. This yields
\begin{align} \label{eq:BSO}
    \delta p_\textrm{e}(t) &= \frac{\Omega_{xz} \cos\vartheta}{2\omegad} \sin\OmegaR t [\sin(\omegad t + \phi_{xz}) - \sin(\phi_{xz})]\\
    &= \frac{\omegaL}{\Omega_y\sin\Delta\phi}\sin\OmegaR t [\sin(\omegad t + \phi_{xz}) - \sin(\phi_{xz})]. \nonumber
\end{align}
The quickly oscillating correction, in the context of resonant drive, is called the Bloch--Siegert oscillation \cite{BlochSiegert1940MNR}, therefore we use the same terminology for the corrections of the RING as well.

The broadening of the “effective resonance” in Fig.~\ref{fig:ring}(c) is captured by the drive frequency range for a fixed Larmor frequency, where the maximum excited state population reaches at least $1/2$. 
The half population inversion for a given Larmor frequency yields the condition $\Delta = \pm\OmegaR$ from Eq.~\eqref{eq:Rabioscillation} in leading order. 
The drive frequency detuning corresponding to this condition, from Eqs.~\eqref{eq:effectiveparameters} and \eqref{eq:ring_freq}, reads as
\begin{eqnarray}
    \Delta \omega^{\pm} = \pm\omega_\textrm{RING} \tan \vartheta.
\end{eqnarray}

The stability of the mechanism can be analyzed through the visibility of the oscillations. From Eq.~\eqref{eq:Rabioscillation}, the visibility is given by
\begin{eqnarray}\label{eq:visibilty}
A = \frac{\OmegaR^2}{\OmegaR^2 + \Delta^2}
=
\frac{1}{1 + \tfrac{\Delta^2}{\OmegaR^2}}.
\end{eqnarray}
Drive imperfections induce deviations in the parameters of the effective Floquet matrix in Eq.~\eqref{eq:Feff} \(\OmegaR\) and \(\Delta\), which we denote by \(\delta\OmegaR\) and \(\delta\Delta\), respectively. Expanding Eq.~\eqref{eq:visibilty} around the effective resonance condition in Eq.~\eqref{eq:effresonance}, the leading-order correction to the visibility reads
\begin{eqnarray}
\delta A = -\bigg(\frac{\delta\Delta}{\OmegaR}\bigg)^2.
\end{eqnarray}
Therefore, the stability of the oscillations against variations of the drive parameters \(\Omega_{xz}\), \(\Omega_{y}\), \(\omegad\), and \(\Delta\phi\) is determined by the induced deviation \(\delta\Delta\).

For a relative error \(\epsilon\) in \(\Omega_{xz}\), \(\Omega_{y}\), or \(\omegad\), one obtains
\begin{eqnarray}
\delta\Delta = \pm\omegaL \epsilon,
\end{eqnarray}
while for a phase error \(\delta\phi\),
\begin{eqnarray}
\delta\Delta = -\omegaL \cot(\Delta\phi)\, \delta\phi.
\end{eqnarray}
The key result of this analysis is that the reduction of the oscillation visibility appears only at second order in the drive errors, moreover choosing $\Delta\phi = \pi/2$ the second-order deviation becomes zero for phase errors. The RING mechanism is therefore intrinsically stable against small imperfections of the drive parameters.

The gyrating nature underlying RING drive is quantified by the factor $\Omega_{xz}\Omega_y \sin\Delta\phi$ in Eqs.~\eqref{eq:effectiveparameters}, which represents the signed area swept by the driving field's elliptical path during each cycle.
This factor changes sign when the handedness of the drive is reversed and vanishes for linearly polarized driving. 
Thus, the RING mechanism requires a gyrating drive field, i.e., a circular or elliptic trajectory in control space with nonzero signed area and definite chirality.
This property of the drive is central to the emergence of asymmetric coupling between Fourier sectors, see Fig.~\ref{fig:floquet}, and hence to the RING mechanism.

\section{RING drive of spin-orbit qubits} \label{sec:ringed}

\begin{figure}[t]
\centering
\includegraphics[width=\linewidth]{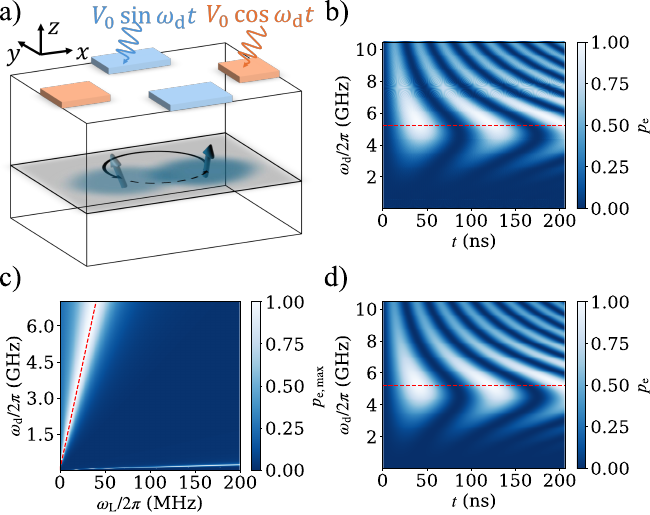}
\caption{Numerical and theoretical demonstration of the RING mechanism in electrically driven spin qubits. (a) Schematic figure of the  electrically driven quantum dot using two electrodes. 
(b) Excited state population as a function of time and drive frequency, from numerical solution of Eq.~\eqref{eq:Hringed}. The parameters are set to the values: $E_0 = 4.4$~V/mm, $\vartheta = \pi/8$, $\alpha=6500$~$\frac{\textrm{m}}{\textrm{s}}$, $\omega_0/2\pi = 140$~GHz, $\omegaL/2\pi = 30$~MHz, $m=0.05m_\textrm{e}$, where $m_\textrm{e}$ is the free electron mass.
Red dashed line: drive frequency corresponding to full Rabi oscillation.
(c) Maximum excited state population as a function of Larmor and drive frequencies, with the same parameters as in (b). Besides the resonant line, full population inversion occurs in the $\omegad \gg \omegaL$ regime. We have plotted the ``effective resonance'' condition in Eq.~\eqref{eq:ringedfreq} with red dashed line.  
(d) Excited state population as a function of time and drive frequency, calculated analytically, with the same parameter values as (b). We marked the frequency of full Rabi oscillations with red dashed line using Eq.~\eqref{eq:ringedfreq}.}
\label{fig:ringed}
\end{figure}

We now consider the RING mechanism in a spin-orbit coupled quantum dot occupied by a single particle with spin-$\tfrac{1}{2}$, where qubit control is achieved via electric fields, see Fig.~\ref{fig:ringed}(a).  
For simplicity, we study an isotropic, harmonic confinement with Rashba spin-orbit coupling, driven by a circularly polarized electric field.
The phenomenon exists in case of a more general confinement potential and spin-orbit interaction as well; we leave the description of such generalized cases for future work.

The Hamiltonian of the real-space effective-mass model of an isotropic quantum dot driven by circularly polarized electric field reads as
\begin{subequations} \label{eq:Hringed}
\begin{align}
    H &= H_0 + \Hso + V_\textrm{d}(t) + \frac{1}{2}\hbar\tilde{\boldsymbol{\omega}}_\mathrm{L}\cdot \boldsymbol{\sigma},\\
    H_0 &= \frac{p^2}{2m} + \frac{1}{2}m\omega_0^2(x^2 + y^2),\\
    \Hso &= \alpha(p_x\sigma_y-p_y\sigma_x),\\
    V_\textrm{d}(t) &= e \mathbf{r}\cdot\mathbf{E}_\textrm{d}(t),
\end{align}
\end{subequations}
where $\hbar\tilde{\boldsymbol{\omega}}$ is the Zeeman field coupled to the spin degree of freedom, in energy units, $m$ is the effective mass of the electron, $\omega_0$ is the strength of the harmonic confinement, $\alpha$ is the strength of the spin-orbit interaction, $e$ is the elementary charge, and
$\mathbf{E}_\textrm{d}(t)$ is a rotating electric field:
\begin{equation}
    \mathbf{E}_\textrm{d}(t) = E_0\begin{pmatrix}
        \cos\omegad t \\
        \sin \omegad t\\
        0
    \end{pmatrix}.
\end{equation}
To avoid a sudden quench at $t=0$, we assume that for $t < 0$ a static electric field is already switched on in the $x$ direction, i.e., $\mathbf{E}_\textrm{d}(t < 0) = \mathbf{E}_\textrm{d}(0)$.

Our goal is to describe the time evolution of an initial state that is in the orbital ground-state subspace, i.e., a superposition of the two lowest-energy eigenstates of $H(t\!=\!0)$. 
Furthermore, we describe the dynamics for an intermediate driving frequency, i.e., when the driving frequency is much greater than the Larmor frequency, but in the limit when 
\begin{equation} \label{eq:adiabaticity}
\omegad e E_0 \ld \ll \hbar \omega_0^2,    
\end{equation}
where $\ld = \sqrt{\hbar/m\omega_0}$ is the confinement length of the quantum dot.
In this case, the probability of the system leaving the orbital ground state is negligible, but significant dynamics is induced within the two lowest-energy instantaneous energy eigenstates of the time-dependent Hamiltonian. 

A natural strategy to describe the dynamics in this case is to compute the time dependence of the instantaneous eigenstates of the time-dependent Hamiltonian, transform the time-dependent Schrodinger equation into that time-dependent instantaneous eigenbasis, and truncate the transformed Hamiltonian matrix to the two-dimensional ground-state subspace. 
Instead of using this approach directly, we convert it to an analytically tractable procedure by a slight modification and a further approximative step. 
The steps of the procedure are as follows.

(1) \textit{Spin-orbit-dressed eigenstates.} We perturbatively compute the energy eigenstates of $H_0 + \Hso$, taking the spin-orbit interaction term as the perturbation. 

(2) \textit{Spatial translation.} We calculate the instantaneous eigenstates of $H_0 + \Hso + V_\textrm{d}(t)$ via the spatial translation of the wavefunctions calculated in the previous step.

(3) \textit{Instantaneous eigenstate frame and truncation.} We transform the time-dependent Schrödinger equation, incorporating the Zeeman term, into the frame defined by these instantaneous, spin-orbit-dressed eigenstates, and truncate the transformed Hamiltonian to the two-dimensional ground-state subspace. 

(4) \textit{Floquet analysis.} Step (3) yields a $2\times 2$ time-dependent effective Hamiltonian. We analyze the induced dynamics using Floquet theory.

\textit{Step (1): Spin-orbit dressed eigenstates.}
As the first step of the derivation of the effective Hamiltonian we calculate the spin-orbit dressed ground-state Kramers pair of $H_0 + \Hso$ using perturbation theory. In the absence of spin-orbit interaction, the ground-state doublet is simply the product of the spin doublet and the ground state of the harmonic confinement:
\begin{eqnarray}
    \ket{\psi_\sigma} &=& \frac{1}{\ld \sqrt{\pi}} e^{-\frac{r^2}{2\ld^2}} \ket{\sigma},
\end{eqnarray}
where $\sigma \in \{\uparrow, \downarrow\}$ labels the eigenstates of $\sigma_z$ and $\ld = \sqrt{\frac{\hbar}{m\omega_0}}$. We apply third-order Schrieffer--Wolff transformation to find the lowest energy subspace of the spin-orbit coupled harmonic oscillator perturbatively \cite{bravyi2011schrieffer, bukov2015universal, winkler2003spinorbit}. The anti-Hermitian generator of the Schrieffer--Wolff transformation reads as
\begin{subequations} \label{eq:SW}
\begin{eqnarray}
    S &=& S^{(1)} + S^{(2)} + S^{(3)},
    \\
    S^{(1)} &=& \frac{i\ld}{\sqrt{2}\lso}(\sigma_x a_y^\dagger   - \sigma_y a_x^\dagger) P_0 - \textrm{H.c.},
    \\
    S^{(2)} &=& -\frac{\ld^2}{4\lso^2}(a_x^{\dagger2} + a_y^{\dagger2})P_0 - \textrm{H.c.},
    \\
    S^{(3)} &=& -\frac{i\ld^3}{12\sqrt{2}\lso^3}
    \bigg[\sigma_x(2a_y^{\dagger} + a_y^{\dagger3} + a_x^{\dagger2}a_y^{\dagger}) 
    \\
    &&\hspace{5em}
    -\sigma_y(2a_x^\dagger + a_x^{\dagger3} + a_y^{\dagger2}a_x^\dagger)\bigg]P_0
    - \textrm{H.c.},\nonumber
\end{eqnarray}
\end{subequations}
where $\lso = \frac{\hbar}{m\alpha}$, $P_0 = \ketbra{\psi_\uparrow} + \ketbra{\psi_\downarrow}$ is the projector to the lowest energy subspace of the 2D isotropic harmonic oscillator without spin-orbit interaction, and $a_x^\dagger$ and $a_y^\dagger$ are creation operators:
\begin{subequations}\begin{eqnarray}
   a_x^\dagger &=& \frac{1}{\sqrt{2}\ld}\bigg(x - \frac{i}{m\omega_0}p_x\bigg),
   \\
   a_y^\dagger &=& \frac{1}{\sqrt{2}\ld}\bigg(y - \frac{i}{m\omega_0}p_y\bigg).   
\end{eqnarray}\end{subequations}
The generators are obtained such that
\begin{eqnarray}
    \tilde H = e^{-S}(H_0 + \Hso)e^{S}
\end{eqnarray}
is block-diagonal up to third order, i.e., the twofold degenerate ground state of $\tilde H$ is decoupled from all other states.
The approximative ground-state Kramers pair of $H_0 + \Hso$ then reads as
\begin{eqnarray} \label{eq:SOdressed}
     \ket{\Psi_\sigma} &=& e^{S} \ket{\psi_\sigma}.
\end{eqnarray}

\textit{Step (2): Spatial translation.} 
The next step is to consider the effect of the driving electric field, which essentially moves the center of the harmonic oscillator on the following trajectory:
\begin{eqnarray} \label{eq:trajectory}
    \mathbf{R}_{\textrm{d}} (t) &=& - \frac{e\mathbf{E}_\textrm{d}(t)}{m\omega_0^2},
\end{eqnarray}
up to a time-dependent constant term in the Hamiltonian. Therefore, to find the instantaneous lowest energy subspace of $H_0 + \Hso + V_\textrm{d}(t)$, we spatially translate the wave function in Eq.~\eqref{eq:SOdressed}. 
The result is a parameter-dependent Schrieffer--Wolff transformation $S_{\mathbf{R}_\textrm{d}}$, where the parameter $\mathbf{R}_\textrm{d}$ only appears in the creation operators
\begin{subequations}
    \label{eq:creation}
\begin{eqnarray}
   a_x^\dagger(\mathbf{R}_\textrm{d}) &=& \frac{1}{\sqrt{2}\ld}\bigg(x - R_{\textrm{d},x}  - \frac{i}{m\omega_0}p_x\bigg),
   \\
   a_y^\dagger(\mathbf{R}_\textrm{d}) &=& \frac{1}{\sqrt{2}\ld}\bigg(y - R_{\textrm{d},y} - \frac{i}{m\omega_0}p_y\bigg),   
\end{eqnarray}
\end{subequations} 
and the projector $P_0(\mathbf{R}_{\textrm{d}})$, which projects to the ground state of the shifted harmonic oscillator:
\begin{eqnarray} \label{eq:shifted_ground}
    \ket{\psi_\sigma (\mathbf{R}_\textrm{d})} = \frac{1}{\ld \sqrt{\pi}} e^{-\frac{(\mathbf{r} - \mathbf{R}_\textrm{d})^2}{2\ld^2}} \ket{\sigma}.
\end{eqnarray}
The instantaneous ground-state doublet of $H_0 + \Hso + V_\textrm{d}(t)$, up to third order, is obtained using Eqs.~\eqref{eq:SW}, \eqref{eq:creation}, and \eqref{eq:shifted_ground}. It  reads as
\begin{widetext}
\begin{multline}\label{eq:Kramers}
    \ket{\Psi_{\sigma} (\mathbf{R}_\textrm{d})} = e^{S_{\mathbf{R}_\textrm{d}}} \ket{\psi_\sigma(\mathbf{R}_\textrm{d})}
    \approx \bigg[\bigg(1 -\frac{\ld^2}{2\lso^2}\bigg)I + \frac{i\ld}{\sqrt{2}\lso}\bigg(1 - \frac{\ld^2}{3\lso^2}\bigg)\big(\sigma_x a_y^\dagger(\mathbf{R}_\textrm{d}) - \sigma_y a_x^\dagger(\mathbf{R}_\textrm{d})\big) - \frac{\ld^2}{4\lso^2}\big(a_x^{\dagger 2}(\mathbf{R}_\textrm{d}) + a_y^{\dagger 2}(\mathbf{R}_\textrm{d})\big)
    \\
    - \frac{i\ld^3}{12\sqrt{2}\lso^3}\bigg(\sigma_x( a_y^{\dagger3}(\mathbf{R}_\textrm{d}) + a_x^{\dagger2}(\mathbf{R}_\textrm{d})a_y^\dagger(\mathbf{R}_\textrm{d}))-\sigma_y(a_x^{\dagger3}(\mathbf{R}_\textrm{d}) + a_x^\dagger(\mathbf{R}_\textrm{d}) a_y^{\dagger2}(\mathbf{R}_\textrm{d}))\bigg)\bigg]\ket{\psi_\sigma(\mathbf{R}_\textrm{d})}.
\end{multline}

\textit{Step (3): Instantaneous eigenstate frame and truncation.} 
The final step to obtain the effective Hamiltonian is to transform the full Hamiltonian to the instantaneous eigenbasis of $H_0 + \Hso + V_\textrm{d}(t)$ and truncate the Hamiltonian to the two-dimensional ground-state subspace.
The result is a $2\times 2$ effective Hamiltonian, that can be separated to a time-independent and a time-dependent part:
\begin{equation} \label{eq:Hq}
    H_\textrm{eff} = \frac{\hbar}{2} (\boldsymbol{\omega}_\textrm{L} + \boldsymbol{\Omega}_\textrm{d}(t))\cdot \boldsymbol{\sigma}.
\end{equation}

The static term is the renormalized Zeeman term:
\begin{eqnarray}
    \boldsymbol{\omega}_\textrm{L}\cdot \boldsymbol{\sigma} &\coloneq& P(\mathbf{R}_\textrm{d})\boldsymbol{\tilde{\omega}}_\textrm{L}\cdot \boldsymbol{\sigma}P(\mathbf{R}_\textrm{d}) = \left(\mathbf{Z}\boldsymbol{\tilde{\omega}}_\textrm{L}\right) \cdot \boldsymbol{\sigma},
\end{eqnarray}
where 
\begin{subequations}\begin{eqnarray}
    P(\mathbf{R}_\textrm{d}) &=& \sum_{\sigma = \uparrow, \downarrow} \ketbra{\Psi_\sigma (\mathbf{R}_\textrm{d})}
\end{eqnarray}\end{subequations}
is the projection to the instantaneous ground-state subspace, and the diagonal matrix 
\begin{subequations}\begin{eqnarray}
    \mathbf{Z} &=& \begin{pmatrix}
        1 - \frac{\ld^2}{\lso^2} & 0 & 0\\
        0 & 1 - \frac{\ld^2}{\lso^2} & 0 \\
        0 & 0 & 1 - \frac{2\ld^2}{\lso^2}
    \end{pmatrix}
\end{eqnarray}\end{subequations}
accounts for the renormalization of the Zeeman term up to third order in $\ld/\lso$ (the third order correction is zero).

The second term is identified as the non-Abelian Berry 
connection \cite{Wilczek1984NABphase, Bohm2003GeometricPhases}, which arises from the time dependence of the eigenstates through the parameters $\mathbf{R}_\textrm{d}(t)$ and reads as
\begin{eqnarray}
    \boldsymbol{\Omega}_\textrm{d}(t)\cdot \boldsymbol{\sigma} &\coloneq& - 2i \sum_{\sigma,\sigma'=\uparrow,\downarrow}\ketbra{\Psi_\sigma(\mathbf{R}_\textrm{d})}{\Psi_{\sigma'}(\mathbf{R}_\textrm{d})} \braket{\Psi_\sigma(\mathbf{R}_\textrm{d})}{ \dot{\mathbf{R}}_\textrm{d}\grad_{\mathbf{R}_\textrm{d}} \Psi_{\sigma'}(\mathbf{R}_\textrm{d})}. \label{eq:BerryConnection}
\end{eqnarray}
To calculate the vector $\boldsymbol{\Omega}_\textrm{d}(t)$ one needs to evaluate the derivative of the basis vectors:
\begin{multline}
    \ket{\partial_{R_{\textrm{d},x}} \Psi_\sigma} = 
    \bigg[\frac{1}{\sqrt{2}\ld}a_x^\dagger 
    + \frac{i}{2\lso}\bigg(1 - \frac{\ld^2}{3\lso^2}\bigg)
    \big[\sigma_y(1-a_x^{\dagger 2}) + \sigma_xa_x^\dagger a_y^\dagger\big] 
    - \frac{\ld}{4\sqrt{2}\lso^2}a_x^{\dagger}\big(a_x^{\dagger 2} + a_y^{\dagger 2}\big)\\
    - \frac{i\ld^2}{24 \lso^3}\bigg(
    \sigma_xa_x^\dagger
    \big[a_y^{\dagger3} + a_x^{\dagger2}a_y^\dagger-
    2a_y^\dagger\big]
    -\sigma_y
    \big[a_x^{\dagger4} 
    + a_x^{\dagger2} a_y^{\dagger2}
    - 3a_x^{\dagger2} 
    - a_y^{\dagger2} 
    \big]\bigg)
    \bigg]\ket{\psi_\sigma},
\end{multline}
\end{widetext} 
where, for brevity, we have omitted the $\mathbf{R}_\textrm{d}$ dependence, and used the identities
\begin{subequations}\begin{eqnarray}
\partial_{R_{\textrm{d},x}}a_x^\dagger(\mathbf{R}_\textrm{d}) &=& -\frac{1}{\sqrt{2}\ld}, \\
\partial_{R_{\textrm{d},x}} \ket{\psi_\sigma(\mathbf{R}_\textrm{d})} &=& \frac{\mathbf{r}-\mathbf{
R_d}}{\ld^2}\ket{\psi_\sigma(\mathbf{R}_\textrm{d})} \\ \nonumber
&=& \frac{1}{\sqrt{2}\ld}a^\dagger_x(\mathbf{R}_\textrm{d})\ket{\psi_\sigma(\mathbf{R}_\textrm{d})}.
\end{eqnarray}\end{subequations}

Finally, multiplying with $\bra{\psi_\sigma(\mathbf{R}_\textrm{d})}$ from the left results in
\begin{equation}\label{eq:dRx}
    \braket{\Psi_\sigma(\mathbf{R}_\textrm{d})}{\partial_{R_{\textrm{d},x}} \Psi_{\sigma'}(\mathbf{R}_\textrm{d})} = 
    \frac{i}{\lso}\bigg(1-\frac{\ld^2}{3\lso^2}\bigg)\bra{\sigma}\sigma_y\ket{\sigma'},
\end{equation}
up to third order in spin-orbit interaction.
Similarly, the other derivative reads as
\begin{equation} \label{eq:dRy}
    \braket{\Psi_\sigma(\mathbf{R}_\textrm{d})}{\partial_{R_{\textrm{d},y}} \Psi_{\sigma'}(\mathbf{R}_\textrm{d})} 
    = -\frac{i}{\lso}\bigg(1-\frac{\ld^2}{3\lso^2}\bigg)\bra{\sigma}\sigma_x\ket{\sigma'}.
\end{equation}
The $\boldsymbol{\Omega}_\textrm{d}(t)$ vector, expressed from  Eq.~\eqref{eq:BerryConnection} using Eqs.~\eqref{eq:trajectory}, \eqref{eq:dRx}, and \eqref{eq:dRy}, reads as 
\begin{eqnarray} \label{eq:effectiveDrive}
    \boldsymbol{\Omega}_\textrm{d}(t) = \frac{2\omegad\tilde{\alpha}  e E_0}{\hbar \omega_0^2}
    \begin{pmatrix}
        \cos\omegad t\\
        \sin\omegad t \\
        0
    \end{pmatrix},
\end{eqnarray}
where $\tilde{\alpha} = \alpha (1 - \ld^2/3\lso^2)$.

\textit{Step (4): Floquet analysis.} The resulting time-dependent two-level Hamiltonian is equivalent to the model in Sec.~\ref{sec:ring}, therefore the same Floquet analysis can be applied. This results in the effective Floquet matrix in Eq.~\eqref{eq:Feff}, with parameters
\begin{subequations}\label{eqs:RashbaRINGED}
\begin{eqnarray}\label{eq:ringed_delta}
    \Delta &=& \omegaL -\frac{2\omegad e^2E_0^2 \tilde{\alpha}^2\cos\vartheta}{\hbar^2\omega_0^4},\\
    \OmegaR &=& \frac{2\omegad e^2E_0^2 \tilde{\alpha}^2\sin\vartheta}{\hbar^2\omega_0^4}, 
\end{eqnarray}
\end{subequations}
where $\vartheta$ is the angle between $\boldsymbol{\omega}_\textrm{L}$ and the $z$ axis. The system performs full Rabi oscillation at the ``effective resonance'', when $\Delta = 0$ in Eq.~\eqref{eq:ringed_delta}. This condition is fulfilled when
\begin{eqnarray}\label{eq:effresonance_ringed}
    \omegaL &=& \frac{2\omegad e^2E_0^2 \tilde{\alpha}^2\cos\vartheta}{\hbar^2\omega_0^4}.
\end{eqnarray}

The drive frequency for full Rabi oscillations, called the RING frequency, is expressed from Eq.~\eqref{eq:effresonance_ringed} as
\begin{eqnarray}\label{eq:ringedfreq}
    \omega_\textrm{RING} = \frac{\hbar^2\omega_0^4}{2 e^2E_0^2 \tilde{\alpha}^2\cos\vartheta} \omegaL.
\end{eqnarray}
The Rabi frequency at the ``effective resonance'' is the same as in Eq.~\eqref{eq:fullRabifreq}, it only depends on the Larmor frequency and the direction of the renormalized Zeeman field $\boldsymbol{\omegaL}$.

We analyzed the dynamics using three complementary methods: (i) full numerical simulation of the real-space Hamiltonian in Eq.~\eqref{eq:Hringed} relying on tight-binding discretization, (ii) a hybrid numerical–analytical solution based on the effective two-level Hamiltonian in Eq.~\eqref{eq:Hq} with $\boldsymbol{\Omega}_\textrm{d}(t)$ given in Eq.~\eqref{eq:effectiveDrive}, and (iii) an analytical solution of the effective Floquet dynamics from Eq.~\eqref{eq:Feff}. 

Figure~\ref{fig:ringed}(b) shows results from method (i).
Similarly to the two-level minimal model presented in the previous section, a chevronlike pattern emerges, with the two differences with respect to the case of resonant drive: The pattern has no symmetry axis at any drive frequency value, and the complete Rabi oscillation appears at a drive frequency different from the drive frequency of the slowest Rabi oscillation.
Figure~\ref{fig:ringed}(c) is obtained with method (ii), and only the maximum value of the excited state probability is plotted. In addition to the conventional resonant response, full population inversion appears at high drive frequency, in agreement with our analysis. 
For comparison, Fig.~\ref{fig:ringed}(d) shows the approximate analytical result from (iii). The plot agrees with the numerical solutions in Fig.~\ref{fig:ringed}(b) up to fine structure, which is the signature of the Bloch--Siegert oscillations, neglected in the analytical solution.

We describe the details of the numerical simulations in the Appendix and we made the code available in \cite{kolok2025ringcode}.

The drive frequency of half population inversion is obtained from the condition $\Delta = \pm \OmegaR$, which results in 
\begin{eqnarray} 
    \Delta \omega^\pm = \frac{\mp\sin\vartheta}{\cos\vartheta \pm\sin\vartheta}\omega_{\textrm{RING}}, 
\end{eqnarray}
where $\omega_{\textrm{RING}}$ is the driving frequency at the ``effective resonance'' defined in Eq.~\eqref{eq:ringedfreq}.

Interestingly, the amplitude of the Bloch--Siegert oscillation does not depend on the drive frequency at the effective resonance, because the effective ac Zeeman field $\boldsymbol{\Omega}_\textrm{d}$ is proportional to $\omegad$. Therefore, from Eqs.~\eqref{eq:BSO} and \eqref{eq:effectiveDrive}, the amplitude reads as
\begin{eqnarray} \label{eq:BSOringed}
    A_\text{BS} = \frac{\tilde{\alpha} e E_0}{\hbar \omega_0^2}\cos\vartheta.
\end{eqnarray}

In this work, we have studied the RING mechanism in the simplest, isotropic Rashba model. 
We note however, that the RING mechanism relies only on the existence of an elliptical effective ac Zeeman field enclosing a finite area in parameter space. 
Therefore, different spin-orbit coupling types and  confinement anisotropy are expected primarily to renormalize the effective ac Zeeman field amplitudes and resonance conditions, rather than qualitatively suppress the mechanism.
We also expect the emergence of the mechanism via $g$-tensor modulation \cite{Crippa2018ElectricalQubits, gyorgy2025gtensorformalism} due to the same reason that it relies only on the existence of an elliptical effective ac Zeeman field enclosing a finite area in parameter space.

\section{Discussion} \label{sec:discussion}

The RING mechanism requires the ability to trace a loop in the qubit control parameter space and strong driving fields with respect to the Larmor frequency, see Figs.~\ref{fig:ring}(c) and \ref{fig:ring}(d). These two key ingredients are achievable in several current quantum hardware platforms. 
Here, we discuss three of those platforms. 

In nitrogen-vacancy center spin qubits, this loop can be realized by applying circularly polarized microwave drives to generate a rotating magnetic field, which has been already demonstrated~\cite{alegre2007polarization, london2014strong}. 
The circular polarization of the driving field is also beneficial due to selection rules: The circularly polarized light only drives one of the transitions between the $m=0$ and $m=\pm1$ spin-1 states. 
Moreover, strong driving amplitudes were also already used to perform coherent control of this platform \cite{fuchs2009gigaherzNV, fuchs2010excited}.

Semiconductor spin qubits, particularly those with strong spin-orbit interaction such as hole-spin qubits in germanium \cite{Hendrickx2021AProcessor, wang2024operating, Hendrickx2024SwwetSensitivity, Froning2021UltrafastFunctionality, wang2024EDSRSihole, Crippa2018ElectricalQubits}, offer another promising platform. 
Here, rotating effective ac Zeeman fields can be generated electrically using multiple gate electrodes, which was already demonstrated in the context of bichromatic driving in Ref.~\cite{john2024bichromatic}. 
In addition, Rabi frequency of the order of 400~MHz with a drive frequency of 3.4~GHz was recently demonstrated \cite{Froning2021UltrafastFunctionality}, which translates to the effective ac Zeeman field strength of 800~MHz. 
This value is already about a factor of two larger than the effective ac Zeeman field strength calculated from the parameter set corresponding to the results shown in Fig.~\ref{fig:ringed} at 3.4 GHz.
We therefore believe that the parameter range required for observing the RING mechanism is compatible with current spin-orbit qubit technology.

Moreover, the Rabi frequency of the RING drive can be at least an order of magnitude larger than the resonant Rabi frequency using the same voltage amplitude at higher frequencies. 
For example, one obtains $\Omega_\textrm{R}/2\pi \approx 0.45~$MHz for the resonant-drive Rabi frequency from the transverse part of the effective ac Zeeman field in Eq.~\eqref{eq:effectiveDrive} with the parameter values: $\alpha = 6500~\tfrac{\textrm{m}}{\textrm{s}}$, $E_0 = 2.5~\tfrac{\textrm{V}}{\textrm{mm}}$, $\omega_0/2\pi = 140~$GHz, $\vartheta = \pi/4$, and $\omegaL/2\pi = \omegad/2\pi = 15~$MHz.
The RING drive induces Rabi oscillations with $\Omega_\textrm{R}/2\pi = \omegaL \tan\vartheta /2\pi = 15~$MHz frequency with the same parameters, but with higher drive frequency: $\omegad/2\pi \approx 10.4~$GHz. This is an amplification by a factor larger than 30. 
Note that the circular resonant drive induces perfect Rabi oscillation, while in the case of RING Bloch--Siegert oscillation emerges with amplitude $A_\textrm{BS} \approx 0.023$, see Eq.~\eqref{eq:BSOringed}.

Superconducting fluxonium qubits also provide a viable platform for RING. 
While the strong drive of flux qubits was demonstrated years ago \cite{Yoshira2014Flux}, a recent work has shown how to simultaneously drive the qubit through both flux and gate voltage, thereby realizing a loop in the qubit’s parameter space~\cite{rower2024suppressing}. 
This dual-drive capability directly enables the gyrating control required for RING.

We also mention potential challenges for quantum logical gates relying on the RING mechanism. 
Bloch--Siegert type oscillations naturally emerge in the RING mechanism, see Eq.~\eqref{eq:BSO}.
One successful mitigation strategy against Bloch--Siegert oscillations is the use of \emph{commensurate pulses}, where the pulse strength, duration, and phase are appropriately chosen to compensate for the effect of Bloch--Siegert oscillations. 
This method has been demonstrated in recent superconducting qubit experiments~\cite{rower2024suppressing}.

Besides Bloch--Siegert oscillations, realistic implementations must also account for finite rise and fall times of the microwave drive, which can in principle lead to adiabatic switching and thereby reduce the visibility of the Rabi oscillations. In recent spin-qubit experiments, rise and fall times in the few-nanosecond range or even below have been used~\cite{fuchs2009gigaherzNV, wang2024operating, van2024coherent}. These timescales are fast compared to the inverse gap of the effective Floquet Hamiltonian, which is typically set by frequencies of a few tens of MHz. We therefore expect adiabatic switching effects to be negligible for such pulses.

\begin{figure}
    \centering
    \includegraphics[width=0.99\linewidth]{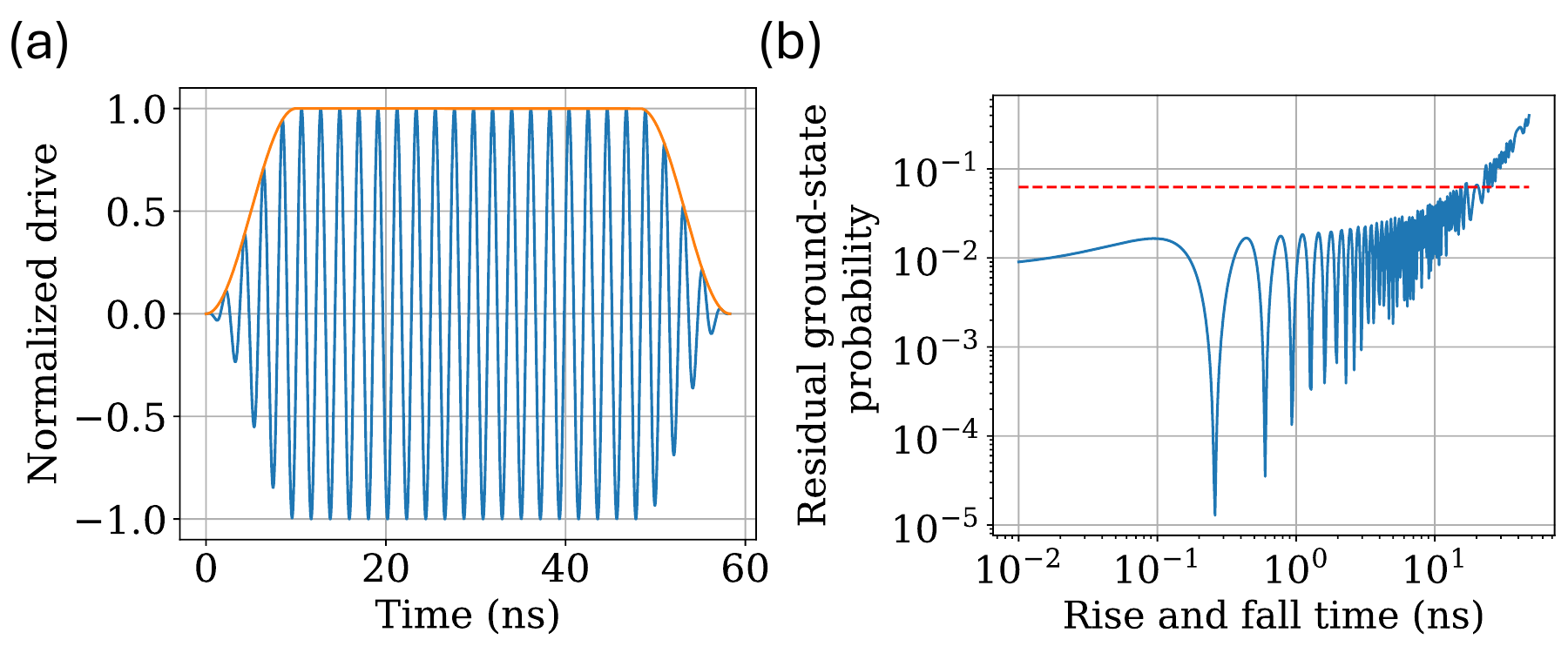}
    \caption{
    Effect of finite rise and fall times on complete population inversion induced by the RING mechanism. 
    (a) Realistic pulse shape, the drive amplitude is multiplied by a smooth envelope with \(\sin^2\)-shaped turn-on and turn-off segments of duration \(t_{\rm r} = 10~\) ns. 
    (b) Residual ground-state probability at half Rabi period as a function of rise time. For each \(t_{\rm r}\), the total pulse duration is chosen such that the integrated envelope area equals that of the corresponding rectangular \(\pi\)-pulse having the same envelope maximum. The plotted quantity is the residual ground-state probability after the pulse. The dashed red line indicates the leading-order micromotion scale \(\omegaL/\Omega_y\), see Eq.~\eqref{eq:BSO}. The nonmonotonic dependence on \(t_{\rm r}\) reflects the phase sensitivity of the residual fast micromotion.
    }
    \label{fig:rise_time}
\end{figure}

To substantiate this expectation, we simulated a RING \(\pi\)-pulse, corresponding to half a Rabi period, with finite rise and fall times. The drive amplitude was multiplied by a smooth envelope with \(\sin^2\)-shaped turn-on and turn-off segments of duration \(t_{\rm r}\). For each value of \(t_{\rm r}\), the total pulse duration was chosen such that the integrated envelope area equals that of the corresponding rectangular \(\pi\)-pulse having the same envelope maximum. 
We emphasize that this area-matching prescription is used only as a simple heuristic for defining a comparable pulse duration. 
Unlike for a conventional resonantly driven two-level system, equal envelope areas do not generally imply equal rotation angles for the RING dynamics. 
The purpose of this calculation is therefore not to determine an optimized shaped $\pi$-pulse, but to demonstrate that substantial population transfer persists even for a smooth, nonoptimized pulse envelope.
The pulse shape is shown in Fig.~\ref{fig:rise_time}(a) and the resulting residual ground-state occupation is shown in Fig.~\ref{fig:rise_time}(b). 
The latter shows that the residual error remains well below the Bloch--Siegert oscillation amplitude, see Eq.~\eqref{eq:BSO}, for nanosecond-scale rise and fall times, demonstrating that realistic smooth envelopes preserve near-complete population inversion in the considered parameter regime.

The dependence on \(t_{\rm r}\) is not strictly monotonic. This is expected because the final error is determined not only by the integrated pulse area, which controls the slow RING rotation, but also by the phase of the fast Bloch--Siegert-type micromotion accumulated during the turn-on, plateau, and turn-off segments. Different ramp times can therefore lead to partial cancellation or enhancement of the residual fast-oscillating correction.

A third challenge is to overcome decoherence, which could wash out the Rabi oscillations.
The coherence time for both semiconductor spin and superconducting qubits is in the microsecond regime or longer \cite{mizuochi2009NVcoherence, Hendrickx2024SwwetSensitivity, stehouwer2025exploiting, rower2024suppressing}.
This is much longer than the typical Rabi period time for the RING mechanism, the latter being in the 100 ns regime, therefore high-visibility Rabi oscillations are expected.

We further emphasize that the RING mechanism highlights several conceptually and practically significant aspects.
First, it constitutes a demonstration of non-Abelian phases in the presence of a nonzero magnetic field, which is of fundamental interest and represents a stepping stone toward holonomic quantum computation. 
Second, since complete Rabi oscillations are achieved under far-detuned driving, the scheme naturally enables the application of high-pass filtering, providing a route to suppress low-frequency and on-resonance noise. This is analogous to subharmonic driving, which permits the suppression of high-frequency and on-resonance noise \cite{schirk2025subharmonicfluxonium}. 
Finally, the mechanism does not rely on resonant energy exchange between the drive and the qubit, in contrast to conventional resonant control protocols.

\section{Conclusions} \label{sec:conclusion}

We have explored an alternative route to coherent qubit control that does not rely on resonant energy exchange. 
Instead, complete Rabi oscillations arise from a distinct mechanism, activated when noncommuting drives trace a finite-area loop in control parameter space. 
With such ingredients, an effective coupling between qubit states emerges from the interference of off-resonant, virtual energy exchange processes.
Using Floquet theory and numerical simulations, we established the conditions under which this nonresonant gyrating drive (RING) produces complete Rabi oscillation in a two-level system. We further demonstrated that the effect can be realized using electric control in spin-orbit coupled quantum dots, leveraging established techniques in qubit manipulation. The required ingredients, including strong, elliptical driving, are accessible in several current experimental platforms, suggesting that this alternative approach to quantum control may be experimentally testable in the near term.

\begin{acknowledgments}
The authors thank G.~Sz\'echenyi, Z.~Gy\"orgy, Bence Het\'enyi, and C.~D\'eprez for useful discussions.
This research was supported by the Ministry of Culture and Innovation (KIM) and the National Research, Development and Innovation Office (NKFIH) within the Quantum Information National Laboratory of Hungary (Grant No.~2022-2.1.1-NL-2022-00004), by the HUN-REN Hungarian Research Network through the HUN-REN-BME-BCE Quantum Technology Research Group, and by the European Union within the Horizon Europe programmes via the projects ONCHIPS (Grant Agreement No.~101080022), IGNITE (Grant Agreement No.~101069515), QLSI2 (Grant Agreement No.~101174557), and OpenSuperQPlus100 (Grant Agreement No.~101113946).
The project supported by the Doctoral Excellence Fellowship Programme (DCEP) is funded by the National Research Development and Innovation Fund of the KIM and the Budapest University of Technology and Economics, under a grant agreement with the NKFIH.
\end{acknowledgments}

\section*{Data availability}
The data that support the findings of this article are openly available~\cite{kolok2025ringcode}.

\appendix

\section*{Appendix: Numerical methods} \label{app:nummeth}
\setcounter{equation}{0}
\renewcommand{\theequation}{A\arabic{equation}}
\renewcommand{\theHequation}{A\arabic{equation}} 

\begin{figure}[t!]
    \centering
    \includegraphics[width=0.95\linewidth]{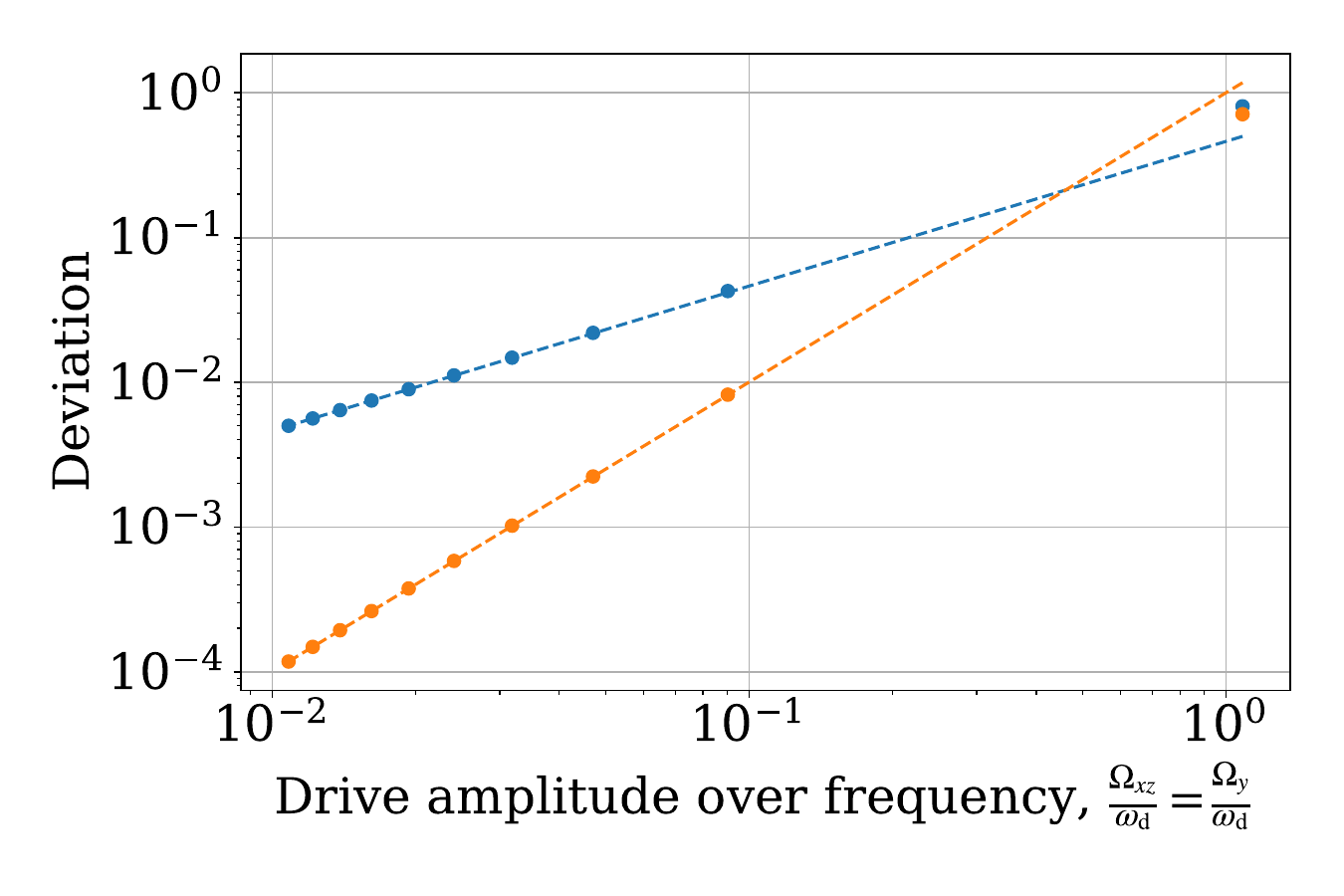}
    \caption{Scaling of the difference between the analytically and numerically determined excited state probabilities.
    The dots are showing the deviation of the numerically calculated excited state probabilities from the analytical, approximate solutions in Eq.~\eqref{eq:Rabioscillation} (blue) and additionally corrected with the formula in Eq.~\eqref{eq:BSO} (orange) as a function of the drive amplitude \(\Omega_{xz} = \Omega_y\).
    The numerical calculation is the solution of the time-dependent Schrödinger equation.
    The drive is circularly polarized and satisfies the effective resonance condition in Eq.~\eqref{eq:effresonance}. Parameters: \(\omegaL/2\pi = 25~\mathrm{MHz},\, \vartheta = \pi/8\).
    The dashed lines show linear (blue) and quadratic (orange) scalings, respectively, normalized to the last numerical data point in each series.}
    \label{fig:errorscaling}
\end{figure}

The numerical solutions shown in Figs.~\ref{fig:ring}(b) and \ref{fig:ring}(c) and Figs.~\ref{fig:ringed}(b) and \ref{fig:ringed}(c) are implemented in Python. 
The time-dependent Schrödinger equation is always solved via the discretization of time. The numerical time-evolution operator reads as
\begin{equation} \label{eq:numericalU}
    U(t) = \prod_{k=0}^{N-1} e^{-\tfrac{i}{\hbar}H(kdt) dt}.
\end{equation}
To obtain the excited state probability we just used Born's rule:
\begin{equation}
    p_\textrm{e}(t) = |\bra{\uparrow}U(t) \ket{\downarrow}|^2.
\end{equation}
To obtain Fig.~\ref{fig:ring}(b), we used $N = 50\,000$ steps. For Fig.~\ref{fig:ringed}(b), we have discretized the Hamiltonian in Eq.~\eqref{eq:Hringed} in space on a $75\times 75$ grid, with $a=5.55~$nm discretization length. 
The dynamics were solved again by using Eq.~\eqref{eq:numericalU}, but at each timestep, we have truncated the Hamiltonian, and only calculated the six smallest eigenvalue-eigenvector pairs to construct the time-evolution operator. The duration of a timestep was $dt = 1/260 \times 2\pi/\omegad$.
To generate each point of Fig.~\ref{fig:ring}(c) and Fig.~\ref{fig:ringed}(c), we have evaluated the numerical time-evolution operator in Eq.~\eqref{eq:numericalU} with the Hamiltonian defined in Eq.~\eqref{eq:Hamiltonian} for Fig.~\ref{fig:ring}(c), and with the effective two-level Hamiltonian in Eq.~\eqref{eq:Hq}, with $\boldsymbol{\Omega}_\textrm{d}(t)$ given in Eq.~\eqref{eq:effectiveDrive}, for Fig.~\ref{fig:ringed}(c). Both the Larmor and driving frequency axes have 400 points. The duration of a timestep was $dt = 1/500\times 2\pi/\omegad$, and the simulations were $T_\textrm{max} = 1000\times 2\pi/\omegad$ long. 

We have validated our numerical results via comparing the simulated excited state probabilities to Eqs.~\eqref{eq:Rabioscillation} and \eqref{eq:BSO} at the effective resonance condition defined by Eq.~\eqref{eq:effresonance} as a function of the \(\frac{\Omega_{xz}}{\omegad} = \frac{\Omega_{y}}{\omegad}\) ratio in Fig.~\ref{fig:errorscaling}. 
The deviation from the leading-order analytical result, see Eq.~\eqref{eq:Rabioscillation}, scales linearly with the perturbation parameter, whereas the deviation from the analytical expression corrected by the formula in Eq.~\eqref{eq:BSO} scales quadratically, as expected. 
This confirms both the validity of the numerical approach and the systematic convergence of the perturbative analytical treatment.

The finite-envelope simulations in Fig.~\ref{fig:rise_time} were performed using the open-source Python package \textsc{Simphony}~\cite{simphony}, which provides tools for simulating spin dynamics and time evolution under user-defined pulse sequences. 
In our simulation, we implemented the Hamiltonian of Eq.~\eqref{eq:Hamiltonian} with time-dependent drive amplitudes multiplied by the \(\sin^2\)-ramp envelope described in Sec.~\ref{sec:discussion}, and extracted the final ground-state occupation after the intended RING \(\pi\)-pulse.

\bibliography{references}

@article{Wilczek1984NABphase,
   author = {Frank Wilczek and A. Zee},
   doi = {10.1103/PhysRevLett.52.2111},
   issn = {0031-9007},
   issue = {24},
   journal = {Phys.~Rev.~Lett.},
   month = {6},
   pages = {2111-2114},
   title = {Appearance of gauge structure in simple dynamical systems},
   volume = {52},
   url = {https://link.aps.org/doi/10.1103/PhysRevLett.52.2111},
   year = {1984}
}

@inbook{Bohm2003GeometricPhases,
author="Bohm, Arno
and Mostafazadeh, Ali
and Koizumi, Hiroyasu
and Niu, Qian
and Zwanziger, Joseph",
title={Mathematical structure of the geometric phase {II}: {The non-Abelian} phase},
bookTitle={The Geometric Phase in Quantum Systems: Foundations, Mathematical Concepts, and Applications in Molecular and Condensed Matter Physics},
year="2003",
publisher="Springer",
address="Berlin, Heidelberg",
pages="129--145",
isbn="978-3-662-10333-3",
doi="10.1007/978-3-662-10333-3_7",
url="https://doi.org/10.1007/978-3-662-10333-3_7"
}

@article{stehouwer2025exploiting,
  title={Exploiting strained epitaxial germanium for scaling low-noise spin qubits at the micrometre scale},
  author={Stehouwer, Lucas E. A. and Yu, C{\'e}cile X and van Straaten, Barnaby and Tosato, Alberto and John, Valentin and Degli Esposti, Davide and Elsayed, Asser and Costa, Davide and Oosterhout, Stefan D and Hendrickx, Nico W and others},
  journal={Nat. Mater.},
  volume={24},
  pages={1906--1912},
  year={2025},
  publisher={Nature Publishing Group UK London},
  doi={10.1038/s41563-025-02276-w}
}

@article{Hendrickx2024SwwetSensitivity,
   author = {N. W. Hendrickx and L. Massai and M. Mergenthaler and F. J. Schupp and S. Paredes and S. W. Bedell and G. Salis and A. Fuhrer},
   doi = {10.1038/s41563-024-01857-5},
   issn = {1476-1122},
   issue = {7},
   journal = {Nat. Mater.},
   month = {7},
   pages = {920-927},
   publisher = {Nature Research},
   title = {Sweet-spot operation of a germanium hole spin qubit with highly anisotropic noise sensitivity},
   volume = {23},
   url = {https://www.nature.com/articles/s41563-024-01857-5},
   year = {2024}
}

@article{zanardi1999holonomic,
  title={Holonomic quantum computation},
  author={Zanardi, P and Rasetti, M},
  journal={Phys. Lett. A},
  volume={264},
  pages={94},
  year={1999},
  doi={10.1016/S0375-9601(99)00803-8}
}

@article{bravyi2011schrieffer,
  title = {{Schrieffer-Wolff} transformation for quantum many-body systems},
  author = {Bravyi, Sergey and DiVincenzo, David P. and Loss, Daniel},
  journal = {Ann. Phys.},
  volume = {326},
  number = {10},
  pages = {2793--2826},
  year = {2011},
  month = {Oct},
  publisher = {Elsevier},
  doi = {10.1016/j.aop.2011.06.004}
}

@article{bukov2015universal,
  title = {Universal high-frequency behavior of periodically driven systems},
  author = {Bukov, Marin and D'Alessio, Luca and Polkovnikov, Anatoli},
  journal = {Adv. Phys.},
  volume = {64},
  number = {2},
  pages = {139--226},
  year = {2015},
  publisher = {Taylor \& Francis},
  doi = {10.1080/00018732.2015.1055918}
}

@article{ballance2016high,
  title={High-fidelity quantum logic gates using trapped-ion hyperfine qubits},
  author={Ballance, C. J. and Harty, T. P. and Linke, N. M. and Sepiol, M. A. and Lucas, D. M.},
  journal={Phys.~Rev.~Lett.},
  volume={117},
  number={6},
  pages={060504},
  year={2016},
  publisher={APS},
  doi={10.1103/PhysRevLett.117.060504}
}

@article{harty2014high,
  title={High-fidelity preparation, gates, memory, and readout of a trapped-ion quantum bit},
  author={Harty, T. P. and Allcock, D. T. C. and Ballance, C. J. and Guidoni, L and Janacek, H. A. and Linke, N. M. and Stacey, D. N. and Lucas, D. M.},
  journal={Phys.~Rev.~Lett.},
  volume={113},
  number={22},
  pages={220501},
  year={2014},
  publisher={APS},
  doi={10.1103/PhysRevLett.113.220501}
}

@article{San-Jose2008GeometricDecoherence,
    title = {Geometric phases in semiconductor spin qubits: {Manipulations} and decoherence},
    year = {2008},
    journal = {Phys.~Rev.~B},
    author = {San-Jose, Pablo and Scharfenberger, Burkhard and Sch{\"{o}}n, Gerd and Shnirman, Alexander and Zarand, Gergely},
    number = {4},
    month = {1},
    pages = {045305},
    volume = {77},
    url = {https://link.aps.org/doi/10.1103/PhysRevB.77.045305},
    doi = {10.1103/PhysRevB.77.045305},
    issn = {1098-0121}
}

@article{mizuochi2009NVcoherence,
  title = {Coherence of single spins coupled to a nuclear spin bath of varying density},
  author = {Mizuochi, N. and Neumann, P. and Rempp, F. and Beck, J. and Jacques, V. and Siyushev, P. and Nakamura, K. and Twitchen, D. J. and Watanabe, H. and Yamasaki, S. and Jelezko, F. and Wrachtrup, J.},
  journal = {Phys. Rev. B},
  volume = {80},
  issue = {4},
  pages = {041201(R)},
  numpages = {4},
  year = {2009},
  month = {Jul},
  publisher = {American Physical Society},
  doi = {10.1103/PhysRevB.80.041201},
  url = {https://link.aps.org/doi/10.1103/PhysRevB.80.041201}
}

@article{unseld2024basebandcontrolsingleelectronsilicon,
  title={Baseband control of single-electron silicon spin qubits in two dimensions},
  author={Unseld, Florian K and Undseth, Brennan and Raymenants, Eline and Matsumoto, Yuta and de Snoo, Sander L and Karwal, Saurabh and Pietx-Casas, Oriol and Ivlev, Alexander S and Meyer, Marcel and Sammak, Amir and others},
  journal={Nat. Commun.},
  volume={16},
  pages={5605},
  year={2025},
  doi={10.1038/s41467-025-60351-x}
}

@article{wang2024operating,
author = {Chien-An Wang  and Valentin John  and Hanifa Tidjani  and Cécile X. Yu  and Alexander S. Ivlev  and Corentin Déprez  and Floor van Riggelen-Doelman  and Benjamin D. Woods  and Nico W. Hendrickx  and William I. L. Lawrie  and Lucas E. A. Stehouwer  and Stefan D. Oosterhout  and Amir Sammak  and Mark Friesen  and Giordano Scappucci  and Sander L. de Snoo  and Maximilian Rimbach-Russ  and Francesco Borsoi  and Menno Veldhorst },
title = {Operating semiconductor quantum processors with hopping spins},
journal = {Science},
volume = {385},
number = {6707},
pages = {447-452},
year = {2024},
doi = {10.1126/science.ado5915}
}

@article{rimbachruss2025gaplessspinqubit,
  title = {Gapless single-spin qubit},
  author = {Rimbach-Russ, Maximilian and John, Valentin and van Straaten, Barnaby and Bosco, Stefano},
  journal = {Phys.~Rev.~Lett.},
  volume = {135},
  issue = {19},
  pages = {197001},
  numpages = {9},
  year = {2025},
  month = {Nov},
  publisher = {American Physical Society},
  doi = {10.1103/mvtj-zhrl},
  url = {https://link.aps.org/doi/10.1103/mvtj-zhrl}
}

@article{john2024bichromatic,
  title = {Bichromatic {Rabi} control of semiconductor qubits},
  author = {John, Valentin and Borsoi, Francesco and Gy\"orgy, Zolt\'an and Wang, Chien-An and Sz\'echenyi, G\'abor and van Riggelen-Doelman, Floor and Lawrie, William I. L. and Hendrickx, Nico W. and Sammak, Amir and Scappucci, Giordano and P\'alyi, Andr\'as and Veldhorst, Menno},
  journal = {Phys.~Rev.~Lett.},
  volume = {132},
  issue = {6},
  pages = {067001},
  numpages = {7},
  year = {2024},
  month = {Feb},
  publisher = {American Physical Society},
  doi = {10.1103/PhysRevLett.132.067001},
  url = {https://link.aps.org/doi/10.1103/PhysRevLett.132.067001}
}

@article{rower2024suppressing,
  title = {Suppressing counter-rotating errors for fast single-qubit gates with fluxonium},
  author = {Rower, David A. and Ding, Leon and Zhang, Helin and Hays, Max and An, Junyoung and Harrington, Patrick M. and Rosen, Ilan T. and Gertler, Jeffrey M. and Hazard, Thomas M. and Niedzielski, Bethany M. and Schwartz, Mollie E. and Gustavsson, Simon and Serniak, Kyle and Grover, Jeffrey A. and Oliver, William D.},
  journal = {PRX Quantum},
  volume = {5},
  issue = {4},
  pages = {040342},
  numpages = {17},
  year = {2024},
  month = {Dec},
  publisher = {American Physical Society},
  doi = {10.1103/PRXQuantum.5.040342},
  url = {https://link.aps.org/doi/10.1103/PRXQuantum.5.040342}
}

@article{alegre2007polarization,
  title = {Polarization-selective excitation of nitrogen vacancy centers in diamond},
  author = {Alegre, Thiago P. Mayer and Santori, Charles and Medeiros-Ribeiro, Gilberto and Beausoleil, Raymond G.},
  journal = {Phys.~Rev.~B},
  volume = {76},
  issue = {16},
  pages = {165205},
  numpages = {5},
  year = {2007},
  month = {Oct},
  publisher = {American Physical Society},
  doi = {10.1103/PhysRevB.76.165205},
  url = {https://link.aps.org/doi/10.1103/PhysRevB.76.165205}
}

@article{london2014strong,
  title = {Strong driving of a single spin using arbitrarily polarized fields},
  author = {London, P. and Balasubramanian, P. and Naydenov, B. and McGuinness, L. P. and Jelezko, F.},
  journal = {Phys.~Rev.~A},
  volume = {90},
  issue = {1},
  pages = {012302},
  numpages = {8},
  year = {2014},
  month = {Jul},
  publisher = {American Physical Society},
  doi = {10.1103/PhysRevA.90.012302},
  url = {https://link.aps.org/doi/10.1103/PhysRevA.90.012302}
}

@article{Yoshira2014Flux,
  title = {Flux qubit noise spectroscopy using {Rabi} oscillations under strong driving conditions},
  author = {Yoshihara, Fumiki and Nakamura, Yasunobu and Yan, Fei and Gustavsson, Simon and Bylander, Jonas and Oliver, William D. and Tsai, Jaw-Shen},
  journal = {Phys.~Rev.~B},
  volume = {89},
  issue = {2},
  pages = {020503(R)},
  numpages = {5},
  year = {2014},
  month = {Jan},
  publisher = {American Physical Society},
  doi = {10.1103/PhysRevB.89.020503},
  url = {https://link.aps.org/doi/10.1103/PhysRevB.89.020503}
}

@article{AbdumalikovJr2013,
   author = {Abdumalikov Jr, A. A.  and Fink, J. M. and Juliusson, K. and Pechal, M. and Berger, S. and Wallraff, A. and Filipp, S.},
   doi = {10.1038/nature12010},
   issn = {0028-0836},
   issue = {7446},
   journal = {Nature},
   month = {4},
   pages = {482-485},
   title = {Experimental realization of {non-Abelian} non-adiabatic geometric gates},
   volume = {496},
   year = {2013}
}

@article{falci2000detection,
  title={Detection of geometric phases in superconducting nanocircuits},
  author={Falci, Giuseppe and Fazio, Rosario and Palma, G Massimo and Siewert, Jens and Vedral, Vlatko},
  journal={Nature},
  volume={407},
  number={6802},
  pages={355--358},
  year={2000},
  publisher={Nature Publishing Group UK London},
  doi={10.1038/35030052}
}

@article{leibfried2003experimental,
  title={Experimental demonstration of a robust, high-fidelity geometric two ion-qubit phase gate},
  author={Leibfried, Dietrich and DeMarco, Brian and Meyer, Volker and Lucas, David and Barrett, Murray and Britton, Joe and Itano, Wayne M and Jelenkovi{\'c}, B and Langer, Chris and Rosenband, Till and others},
  journal={Nature},
  volume={422},
  number={6930},
  pages={412--415},
  year={2003},
  publisher={Nature Publishing Group UK London},
  doi={10.1038/nature01492}
}

@article{toyoda2013realization,
  title={Realization of holonomic single-qubit operations},
  author={Toyoda, K and Uchida, K and Noguchi, A and Haze, S and Urabe, S},
  journal={Phys.~Rev.~A},
  volume={87},
  number={5},
  pages={052307},
  year={2013},
  publisher={APS},
  doi={10.1103/PhysRevA.87.052307}
}

@article{arroyo2014room,
  title={Room temperature high-fidelity holonomic single-qubit gate on a solid-state spin},
  author={Arroyo-Camejo, Silvia and Lazariev, Andrii and Hell, Stefan W and Balasubramanian, Gopalakrishnan},
  journal={Nat. Commun.},
  volume={5},
  number={1},
  pages={4870},
  year={2014},
  publisher={Nature Publishing Group UK London},
  doi={10.1038/ncomms5870}
}

@article{zu2014experimental,
  title={Experimental realization of universal geometric quantum gates with solid-state spins},
  author={Zu, Chong and Wang, W-B and He, Li and Zhang, W-G and Dai, C-Y and Wang, Fei and Duan, L-M},
  journal={Nature},
  volume={514},
  number={7520},
  pages={72--75},
  year={2014},
  publisher={Nature Publishing Group UK London},
  doi={10.1038/nature13729}
}

@article{yale2016optical,
  author  = {Yale, C. G. and Heremans, F. J. and Zhou, B. B. and Auer, A. and Burkard, G. and Awschalom, D. D.},
  title   = {Optical manipulation of the {Berry} phase in a solid-state spin qubit},
  journal = {Nat. Photon.},
  volume  = {10},
  pages   = {184},
  year    = {2016}
}

@article{sekiguchi2017optical,
  title={Optical holonomic single quantum gates with a geometric spin under a zero field},
  author={Sekiguchi, Yuhei and Niikura, Naeko and Kuroiwa, Ryota and Kano, Hiroki and Kosaka, Hideo},
  journal={Nat. Photon.},
  volume={11},
  number={5},
  pages={309--314},
  year={2017},
  publisher={Nature Publishing Group UK London},
  doi={10.1038/nphoton.2017.40}
}

@article{ishida2018universal,
  title={Universal holonomic single quantum gates over a geometric spin with phase-modulated polarized light},
  author={Ishida, Naoki and Nakamura, Takaaki and Tanaka, Touta and Mishima, Shota and Kano, Hiroki and Kuroiwa, Ryota and Sekiguchi, Yuhei and Kosaka, Hideo},
  journal={Opt. Lett.},
  volume={43},
  number={10},
  pages={2380--2383},
  year={2018},
  publisher={Optical Society of America},
  doi={10.1364/OL.43.002380}
}

@article{zhou2017holonomic,
  title={Holonomic quantum control by coherent optical excitation in diamond},
  author={Zhou, Brian B and Jerger, Paul C and Shkolnikov, Vladislav O and Heremans, F Joseph and Burkard, Guido and Awschalom, David D},
  journal={Phys.~Rev.~Lett.},
  volume={119},
  number={14},
  pages={140503},
  year={2017},
  publisher={APS},
  doi={10.1103/PhysRevLett.119.140503}
}

@article{nagata2018universal,
  title={Universal holonomic quantum gates over geometric spin qubits with polarised microwaves},
  author={Nagata, Kodai and Kuramitani, Kouyou and Sekiguchi, Yuhei and Kosaka, Hideo},
  journal={Nat. Commun.},
  volume={9},
  number={1},
  pages={3227},
  year={2018},
  publisher={Nature Publishing Group UK London},
  doi={10.1038/s41467-018-05664-w}
}

@article{hong2018implementing,
  title={Implementing universal nonadiabatic holonomic quantum gates with transmons},
  author={Hong, Zhuo-Ping and Liu, Bao-Jie and Cai, Jia-Qi and Zhang, Xin-Ding and Hu, Yong and Wang, Z. D. and Xue, Zheng-Yuan},
  journal={Phys.~Rev.~A},
  volume={97},
  number={2},
  pages={022332},
  year={2018},
  publisher={APS},
  doi={10.1103/PhysRevA.97.022332}
}

@article{kjaergaard2020superconducting,
  title={Superconducting qubits: {Current} state of play},
  author={Kjaergaard, Morten and Schwartz, Mollie E and Braum{\"u}ller, Jochen and Krantz, Philip and Wang, Joel I-J and Gustavsson, Simon and Oliver, William D},
  journal={Annu. Rev. Condens. Matter Phys.},
  volume={11},
  number={1},
  pages={369--395},
  year={2020},
  publisher={Annual Reviews},
  doi={10.1146/annurev-conmatphys-031119-050605}
}

@article{Nakamura1999CoherentBox,
    title = {Coherent control of macroscopic quantum states in a {single-Cooper-pair} box},
    year = {1999},
    journal = {Nature},
    author = {Nakamura, Y. and Pashkin, Yu. A. and Tsai, J. S.},
    number = {6730},
    month = {4},
    pages = {786--788},
    volume = {398},
    doi = {10.1038/19718},
    issn = {0028-0836}
}

@article{Vion2002ManipulatingCircuit,
    title = {Manipulating the quantum state of an electrical circuit},
    year = {2002},
    journal = {Science},
    author = {Vion, D. and Aassime, A. and Cottet, A. and Joyez, P. and Pothier, H. and Urbina, C. and Esteve, D. and Devoret, M. H.},
    number = {5569},
    month = {5},
    pages = {886--889},
    volume = {296},
    doi = {10.1126/science.1069372},
    issn = {0036-8075}
}

@article{chiorescu2003coherent,
  title={Coherent quantum dynamics of a superconducting flux qubit},
  author={Chiorescu, Irinel and Nakamura, Y and Harmans, CJP Ma and Mooij, JE},
  journal={Science},
  volume={299},
  number={5614},
  pages={1869--1871},
  year={2003},
  publisher={American Association for the Advancement of Science},
  doi={10.1126/science.1081045}
}

@article{wallraff2004strong,
  title={Strong coupling of a single photon to a superconducting qubit using circuit quantum electrodynamics},
  author={Wallraff, Andreas and Schuster, David I and Blais, Alexandre and Frunzio, Luigi and Huang, R-S and Majer, Johannes and Kumar, Sameer and Girvin, Steven M and Schoelkopf, Robert J},
  journal={Nature},
  volume={431},
  number={7005},
  pages={162--167},
  year={2004},
  publisher={Nature Publishing Group UK London},
  doi={10.1038/nature02851}
}

@article{manucharyan2009fluxonium,
  title={Fluxonium: {Single} {Cooper-pair} circuit free of charge offsets},
  author={Manucharyan, Vladimir E and Koch, Jens and Glazman, Leonid I and Devoret, Michel H},
  journal={Science},
  volume={326},
  number={5949},
  pages={113--116},
  year={2009},
  publisher={American Association for the Advancement of Science},
  doi={10.1126/science.1175552}
}

@article{van2024coherent,
  title={Coherent spin qubit shuttling through germanium quantum dots},
  author={van Riggelen-Doelman, Floor and Wang, Chien-An and de Snoo, Sander L and Lawrie, William I. L. and Hendrickx, Nico W and Rimbach-Russ, Maximilian and Sammak, Amir and Scappucci, Giordano and D{\'e}prez, Corentin and Veldhorst, Menno},
  journal={Nat. Commun.},
  volume={15},
  number={1},
  pages={5716},
  year={2024},
  publisher={Nature Publishing Group UK London},
  doi={10.1038/s41467-024-49358-y}
}

@article{romhanyi2015subharmonic,
   author = {Judit Romhányi and Guido Burkard and András Pályi},
   doi = {10.1103/PhysRevB.92.054422},
   issn = {1098-0121},
   issue = {5},
   journal = {Phys.~Rev.~B},
   month = {8},
   pages = {054422},
   title = {Subharmonic transitions and {Bloch-Siegert} shift in electrically driven spin resonance},
   volume = {92},
   url = {https://link.aps.org/doi/10.1103/PhysRevB.92.054422},
   year = {2015}
}

@article{shirley1965floquet,
  title = {Solution of the {Schr\"odinger} equation with a {Hamiltonian} periodic in time},
  author = {Shirley, Jon H.},
  journal = {Phys.~Rev.},
  volume = {138},
  issue = {4B},
  pages = {B979--B987},
  numpages = {0},
  year = {1965},
  month = {May},
  publisher = {American Physical Society},
  doi = {10.1103/PhysRev.138.B979},
  url = {https://link.aps.org/doi/10.1103/PhysRev.138.B979}
}

@Inbook{winkler2003spinorbit,
author="Winkler, Roland",
title     = {Quasi-degenerate perturbation theory},
booktitle = {Spin-Orbit Coupling Effects in Two-Dimensional Electron and Hole Systems},
year="2003",
publisher="Springer",
address="Berlin, Heidelberg",
pages="201--205",
isbn="978-3-540-36616-4",
doi="10.1007/978-3-540-36616-4_12",
url="https://doi.org/10.1007/978-3-540-36616-4_12"
}

@article{kolok2024protocols,
  title = {Protocols to measure the {non-Abelian Berry} phase by pumping a spin qubit through a quantum-dot loop},
  author = {Kolok, Baksa and P\'alyi, Andr\'as},
  journal = {Phys.~Rev.~B},
  volume = {109},
  issue = {4},
  pages = {045438},
  numpages = {16},
  year = {2024},
  month = {Jan},
  publisher = {American Physical Society},
  doi = {10.1103/PhysRevB.109.045438},
  url = {https://link.aps.org/doi/10.1103/PhysRevB.109.045438}
}

@article{golovach2010holonomicspinqubits,
  title = {Holonomic quantum computation with electron spins in quantum dots},
  author = {Golovach, Vitaly N. and Borhani, Massoud and Loss, Daniel},
  journal = {Phys.~Rev.~A},
  volume = {81},
  issue = {2},
  pages = {022315},
  numpages = {7},
  year = {2010},
  month = {Feb},
  publisher = {American Physical Society},
  doi = {10.1103/PhysRevA.81.022315},
  url = {https://link.aps.org/doi/10.1103/PhysRevA.81.022315}
}

@misc{kolok2025ringcode,
  author       = {Kolok, Baksa},
  title        = {Code to generate figures for ``{Rabi} oscillations induced by nonresonant gyrating drive''},
  howpublished = {Zenodo},
  note = {[Computer software], \url{https://doi.org/10.5281/zenodo.23156356}},
  year         = {2026},
  doi          = {10.5281/zenodo.23156356},
  url          = {https://doi.org/10.5281/zenodo.23156356}
}

@article{Levy2002Singlettriplettheory,
  title = {Universal quantum computation with spin-$1/2$ pairs and {Heisenberg} exchange},
  author = {Levy, Jeremy},
  journal = {Phys.~Rev.~Lett.},
  volume = {89},
  issue = {14},
  pages = {147902},
  numpages = {3},
  year = {2002},
  month = {Sep},
  publisher = {American Physical Society},
  doi = {10.1103/PhysRevLett.89.147902},
  url = {https://link.aps.org/doi/10.1103/PhysRevLett.89.147902}
}

@article{Petta2005singlettripletfirst,
author = {J. R. Petta  and A. C. Johnson  and J. M. Taylor  and E. A. Laird  and A. Yacoby  and M. D. Lukin  and C. M. Marcus  and M. P. Hanson  and A. C. Gossard },
title = {Coherent manipulation of coupled electron spins in semiconductor quantum dots},
journal = {Science},
volume = {309},
number = {5744},
pages = {2180-2184},
year = {2005},
doi = {10.1126/science.1116955},
URL = {https://www.science.org/doi/abs/10.1126/science.1116955}}

@article{Maune2012singlettriplet,
  title={Coherent singlet-triplet oscillations in a silicon-based double quantum dot},
  author={Maune, Brett M and Borselli, Matthew G and Huang, Biqin and Ladd, Thaddeus D and Deelman, Peter W and Holabird, Kevin S and Kiselev, Andrey A and Alvarado-Rodriguez, Ivan and Ross, Richard S and Schmitz, Adele E and others},
  journal={Nature},
  volume={481},
  number={7381},
  pages={344--347},
  year={2012},
  publisher={Nature Publishing Group UK London},
  doi={10.1038/nature10707}
}

@article{Wu2014twoaxiscontrol,
  title={Two-axis control of a singlet--triplet qubit with an integrated micromagnet},
  author={Wu, Xian and Ward, Daniel R and Prance, JR and Kim, Dohun and Gamble, John King and Mohr, RT and Shi, Zhan and Savage, DE and Lagally, MG and Friesen, Mark and others},
  journal={Proc. Natl. Acad. Sci. U.S.A.},
  volume={111},
  number={33},
  pages={11938--11942},
  year={2014},
  publisher={National Academy of Sciences},
  doi={10.1073/pnas.1412230111}
}

@article{Jirovec2021singlettripletGe,
  title={A singlet-triplet hole spin qubit in planar {Ge}},
  author={Jirovec, Daniel and Hofmann, Andrea and Ballabio, Andrea and Mutter, Philipp M and Tavani, Giulio and Botifoll, Marc and Crippa, Alessandro and Kukucka, Josip and Sagi, Oliver and Martins, Frederico and others},
  journal={Nat. Mater.},
  volume={20},
  number={8},
  pages={1106--1112},
  year={2021},
  publisher={Nature Publishing Group UK London},
  doi={10.1038/s41563-021-01022-2}
}

@article{Zhang2025foursinglettriplet,
  title={Universal control of four singlet--triplet qubits},
  author={Zhang, Xin and Morozova, Elizaveta and Rimbach-Russ, Maximilian and Jirovec, Daniel and Hsiao, Tzu-Kan and Fari{\~n}a, Pablo Cova and Wang, Chien-An and Oosterhout, Stefan D and Sammak, Amir and Scappucci, Giordano and others},
  journal={Nat. Nanotechnol.},
  volume={20},
  number={2},
  pages={209--215},
  year={2025},
  publisher={Nature Publishing Group UK London},
  doi={10.1038/s41565-024-01817-9}
}

@article{Kempe2001exchangeonlytheory2,
  title={Theory of decoherence-free fault-tolerant universal quantum computation},
  author={Kempe, Julia and Bacon, Dave and Lidar, Daniel A and Whaley, K Birgitta},
  journal={Phys.~Rev.~A},
  volume={63},
  number={4},
  pages={042307},
  year={2001},
  publisher={APS},
  doi={10.1103/PhysRevA.63.042307}
}

@article{Divincenzo2000exchangeonlytheory1,
  title={Universal quantum computation with the exchange interaction},
  author={DiVincenzo, David P and Bacon, Dave and Kempe, Julia and Burkard, Guido and Whaley, K Birgitta},
  journal={Nature},
  volume={408},
  number={6810},
  pages={339--342},
  year={2000},
  publisher={Nature Publishing Group UK London},
  doi={10.1038/35042541}
}

@article{Weinstein2023exchangeonlyexp,
  title={Universal logic with encoded spin qubits in silicon},
  author={Weinstein, Aaron J and Reed, Matthew D and Jones, Aaron M and Andrews, Reed W and Barnes, David and Blumoff, Jacob Z and Euliss, Larken E and Eng, Kevin and Fong, Bryan H and Ha, Sieu D and others},
  journal={Nature},
  volume={615},
  number={7954},
  pages={817--822},
  year={2023},
  publisher={Nature Publishing Group UK London},
  doi={10.1038/s41586-023-05777-3}
}

@article{sjoqvist2012nonadiabaticholonomic,
  title={Non-adiabatic holonomic quantum computation},
  author={Sj{\"o}qvist, Erik and Tong, Dian-Min and Mauritz Andersson, L and Hessmo, Bj{\"o}rn and Johansson, Markus and Singh, Kuldip},
  journal={New J. Phys.},
  volume={14},
  number={10},
  pages={103035},
  year={2012},
  publisher={IOP Publishing},
  doi={10.1088/1367-2630/14/10/103035}
}

@article{
fuchs2009gigaherzNV,
author = {G. D. Fuchs  and V. V. Dobrovitski  and D. M. Toyli  and F. J. Heremans  and D. D. Awschalom },
title = {Gigahertz Dynamics of a Strongly Driven Single Quantum Spin},
journal = {Science},
volume = {326},
number = {5959},
pages = {1520-1522},
year = {2009},
doi = {10.1126/science.1181193},
URL = {https://www.science.org/doi/abs/10.1126/science.1181193}
}

@article{fuchs2010excited,
  title={Excited-state spin coherence of a single nitrogen--vacancy centre in diamond},
  author={Fuchs, GD and Dobrovitski, VV and Toyli, DM and Heremans, FJ and Weis, CD and Schenkel, Thomas and Awschalom, DD},
  journal={Nat. Phys.},
  volume={6},
  number={9},
  pages={668--672},
  year={2010},
  publisher={Nature Publishing Group UK London},
  doi={10.1038/nphys1716}
}

@article{Bosco2024exchangeonlyspinorbitqubitssilicon,
  title = {Exchange-only spin-orbit qubits in silicon and germanium},
  author = {Bosco, Stefano and Rimbach-Russ, Maximilian},
  journal = {Phys. Rev. Appl.},
  volume = {25},
  issue = {2},
  pages = {L021002},
  numpages = {6},
  year = {2026},
  month = {Feb},
  publisher = {American Physical Society},
  doi = {10.1103/nl35-6886},
  url = {https://link.aps.org/doi/10.1103/nl35-6886}
}

@article{Schrader2004NeutralAtoms,
  title = {Neutral Atom Quantum Register},
  author = {Schrader, D. and Dotsenko, I. and Khudaverdyan, M. and Miroshnychenko, Y. and Rauschenbeutel, A. and Meschede, D.},
  journal = {Phys. Rev. Lett.},
  volume = {93},
  issue = {15},
  pages = {150501},
  numpages = {4},
  year = {2004},
  month = {Oct},
  publisher = {American Physical Society},
  doi = {10.1103/PhysRevLett.93.150501},
  url = {https://link.aps.org/doi/10.1103/PhysRevLett.93.150501}
}

@article{Wang2015NeutralAtomIndividual,
  title = {Coherent Addressing of Individual Neutral Atoms in a {3D} Optical Lattice},
  author = {Wang, Yang and Zhang, Xianli and Corcovilos, Theodore A. and Kumar, Aishwarya and Weiss, David S.},
  journal = {Phys.~Rev.~Lett.},
  volume = {115},
  issue = {4},
  pages = {043003},
  numpages = {5},
  year = {2015},
  month = {Jul},
  publisher = {American Physical Society},
  doi = {10.1103/PhysRevLett.115.043003},
  url = {https://link.aps.org/doi/10.1103/PhysRevLett.115.043003}
}

@article{Xia2015NeutralAtomsRandBench,
  title = {Randomized Benchmarking of Single-Qubit Gates in a {2D} Array of Neutral-Atom Qubits},
  author = {Xia, T. and Lichtman, M. and Maller, K. and Carr, A. W. and Piotrowicz, M. J. and Isenhower, L. and Saffman, M.},
  journal = {Phys. Rev. Lett.},
  volume = {114},
  issue = {10},
  pages = {100503},
  numpages = {5},
  year = {2015},
  month = {Mar},
  publisher = {American Physical Society},
  doi = {10.1103/PhysRevLett.114.100503},
  url = {https://link.aps.org/doi/10.1103/PhysRevLett.114.100503}
}

@article{BlochSiegert1940MNR,
  title = {Magnetic Resonance for Nonrotating Fields},
  author = {Bloch, F. and Siegert, A.},
  journal = {Phys.~Rev.},
  volume = {57},
  issue = {6},
  pages = {522--527},
  numpages = {0},
  year = {1940},
  month = {Mar},
  publisher = {American Physical Society},
  doi = {10.1103/PhysRev.57.522},
  url = {https://link.aps.org/doi/10.1103/PhysRev.57.522}
}

@article{takahiro2016BWtheory,
  title = {{Brillouin-Wigner} theory for high-frequency expansion in periodically driven systems: Application to {Floquet} topological insulators},
  author = {Mikami, Takahiro and Kitamura, Sota and Yasuda, Kenji and Tsuji, Naoto and Oka, Takashi and Aoki, Hideo},
  journal = {Phys. Rev. B},
  volume = {93},
  issue = {14},
  pages = {144307},
  numpages = {25},
  year = {2016},
  month = {Apr},
  publisher = {American Physical Society},
  doi = {10.1103/PhysRevB.93.144307},
  url = {https://link.aps.org/doi/10.1103/PhysRevB.93.144307}
}

@article{saar2003effectivehamiltonian,
  title = {Effective {Hamiltonians} for periodically driven systems},
  author = {Rahav, Saar and Gilary, Ido and Fishman, Shmuel},
  journal = {Phys. Rev. A},
  volume = {68},
  issue = {1},
  pages = {013820},
  numpages = {18},
  year = {2003},
  month = {Jul},
  publisher = {American Physical Society},
  doi = {10.1103/PhysRevA.68.013820},
  url = {https://link.aps.org/doi/10.1103/PhysRevA.68.013820}
}

@article{Hendrickx2021AProcessor,
    title = {{A four-qubit germanium quantum processor}},
    year = {2021},
    journal = {Nature},
    author = {Hendrickx, Nico W. and Lawrie, William I. L. and Russ, Maximilian and van Riggelen, Floor and de Snoo, Sander L. and Schouten, Raymond N. and Sammak, Amir and Scappucci, Giordano and Veldhorst, Menno},
    number = {7851},
    month = {3},
    pages = {580--585},
    volume = {591},
    publisher = {Nature Research},
    url = {https://www.nature.com/articles/s41586-021-03332-6},
    doi = {10.1038/s41586-021-03332-6},
    issn = {0028-0836}
}

@article{gyorgy2025gtensorformalism,
  title = {Limitations of the $g$-tensor formalism of semiconductor spin qubits},
  author = {Gy\"orgy, Zolt\'an and P\'alyi, Andr\'as and Sz\'echenyi, G\'abor},
  journal = {Phys. Rev. B},
  volume = {112},
  issue = {4},
  pages = {045428},
  numpages = {18},
  year = {2025},
  month = {Jul},
  publisher = {American Physical Society},
  doi = {10.1103/gq3h-76mx},
  url = {https://link.aps.org/doi/10.1103/gq3h-76mx}
}

@article{Pla2012ASilicon,
    title = {{A single-atom electron spin qubit in silicon}},
    year = {2012},
    journal = {Nature},
    author = {Pla, Jarryd J. and Tan, Kuan Y. and Dehollain, Juan P. and Lim, Wee H. and Morton, John J. L. and Jamieson, David N. and Dzurak, Andrew S. and Morello, Andrea},
    number = {7417},
    month = {9},
    pages = {541--545},
    volume = {489},
    doi = {10.1038/nature11449},
    issn = {0028-0836}
}

@article{Hile2018AddressableSilicon,
    title = {{Addressable electron spin resonance using donors and donor molecules in silicon}},
    year = {2018},
    journal = {Sci. Adv.},
    author = {Hile, Samuel J. and Fricke, Lukas and House, Matthew G. and Peretz, Eldad and Chen, Chin Yi and Wang, Yu and Broome, Matthew and Gorman, Samuel K. and Keizer, Joris G. and Rahman, Rajib and Simmons, Michelle Y.},
    number = {7},
    month = {7},
    pages = {eaaq1459},
    volume = {4},
    url = {https://www.science.org/doi/10.1126/sciadv.aaq1459},
    doi = {10.1126/sciadv.aaq1459},
    issn = {2375-2548}
}

@article{Veldhorst2014AnControl-fidelity,
    title = {{An addressable quantum dot qubit with fault-tolerant control-fidelity}},
    year = {2014},
    journal = {Nat. Nanotechnol.},
    author = {Veldhorst, M. and Hwang, J. C. C. and Yang, C. H. and Leenstra, A. W. and de Ronde, B. and Dehollain, J. P. and Muhonen, J. T. and Hudson, F. E. and Itoh, K. M. and Morello, A. and Dzurak, A. S.},
    number = {12},
    month = {12},
    pages = {981--985},
    volume = {9},
    doi = {10.1038/nnano.2014.216},
    issn = {1748-3387}
}

@article{Nowack2007CoherentFields,
    title = {{Coherent control of a single electron spin with electric fields}},
    year = {2007},
    journal = {Science},
    author = {Nowack, K. C. and Koppens, F. H. L. and Nazarov, Yu. V. and Vandersypen, L. M. K.},
    number = {5855},
    month = {11},
    pages = {1430--1433},
    volume = {318},
    url = {https://www.science.org/doi/10.1126/science.1148092},
    doi = {10.1126/science.1148092},
    issn = {0036-8075}
}

@article{Koppens2006DrivenDot,
    title = {{Driven coherent oscillations of a single electron spin in a quantum dot}},
    year = {2006},
    journal = {Nature},
    author = {Koppens, F. H.L. and Buizert, C. and Tielrooij, K. J. and Vink, I. T. and Nowack, K. C. and Meunier, T. and Kouwenhoven, L. P. and Vandersypen, L. M.K.},
    number = {7104},
    month = {8},
    pages = {766--771},
    volume = {442},
    publisher = {Nature Publishing Group},
    doi = {10.1038/nature05065},
    issn = {14764687}
}

@article{Crippa2018ElectricalQubits,
  author  = {Crippa, A. and Maurand, R. and Bourdet, L. and Kotekar-Patil, D. and Amisse, A. and Jehl, X. and Sanquer, M. and Lavi{\'e}ville, R. and Bohuslavskyi, H. and Hutin, L. and Barraud, S. and Vinet, M. and Niquet, Y. M. and De Franceschi, S.},
  title   = {Electrical spin driving by $g$-matrix modulation in spin-orbit qubits},
  journal = {Phys. Rev. Lett.},
  volume  = {120},
  pages   = {137702},
  year    = {2018}
}

@article{Pla2013High-fidelitySilicon,
    title = {{High-fidelity readout and control of a nuclear spin qubit in silicon}},
    year = {2013},
    journal = {Nature},
    author = {Pla, Jarryd J. and Tan, Kuan Y. and Dehollain, Juan P. and Lim, Wee H. and Morton, John J. L. and Zwanenburg, Floris A. and Jamieson, David N. and Dzurak, Andrew S. and Morello, Andrea},
    number = {7445},
    month = {4},
    pages = {334--338},
    volume = {496},
    doi = {10.1038/nature12011},
    issn = {0028-0836}
}

@article{Tettamanzi2017ProbingFrequencies,
    title = {{Probing the quantum states of a single atom transistor at microwave frequencies}},
    year = {2017},
    journal = {ACS Nano},
    author = {Tettamanzi, Giuseppe Carlo and Hile, Samuel James and House, Matthew Gregory and Fuechsle, Martin and Rogge, Sven and Simmons, Michelle Y.},
    number = {3},
    month = {3},
    pages = {2444--2451},
    volume = {11},
    doi = {10.1021/acsnano.6b06362},
    issn = {1936-0851}
}

@article{Wang2024PursuingDot,
    title = {{Pursuing high-fidelity control of spin qubits in natural Si/SiGe quantum dot}},
    year = {2024},
    journal = {Appl. Phys. Lett.},
    author = {Wang, Ning and Wang, Shao-Min and Zhang, Run-Ze and Kang, Jia-Min and Lu, Wen-Long and Li, Hai-Ou and Cao, Gang and Wang, Bao-Chuan and Guo, Guo-Ping},
    number = {20},
    month = {11},
    pages = {204002},
    volume = {125},
    url = {https://pubs.aip.org/apl/article/125/20/204002/3319548/Pursuing-high-fidelity-control-of-spin-qubits-in},
    doi = {10.1063/5.0230605},
    issn = {0003-6951}
}

@article{Burkard2023SemiconductorQubits,
    title = {{Semiconductor spin qubits}},
    year = {2023},
    journal = {Rev. Mod. Phys.},
    author = {Burkard, Guido and Ladd, Thaddeus D. and Pan, Andrew and Nichol, John M. and Petta, Jason R.},
    number = {2},
    month = {6},
    pages = {025003},
    volume = {95},
    publisher = {American Physical Society},
    url = {https://link.aps.org/doi/10.1103/RevModPhys.95.025003},
    doi = {10.1103/RevModPhys.95.025003},
    issn = {0034-6861}
}

@article{Koch2019SpinTransistor,
    title = {{Spin read-out in atomic qubits in an all-epitaxial three-dimensional transistor}},
    year = {2019},
    journal = {Nat. Nanotechnol.},
    author = {Koch, Matthias and Keizer, Joris G. and Pakkiam, Prasanna and Keith, Daniel and House, Matthew G. and Peretz, Eldad and Simmons, Michelle Y.},
    number = {2},
    month = {2},
    pages = {137--140},
    volume = {14},
    doi = {10.1038/s41565-018-0338-1},
    issn = {1748-3387}
}

@article{Froning2021UltrafastFunctionality,
    title = {{Ultrafast hole spin qubit with gate-tunable spin–orbit switch functionality}},
    year = {2021},
    journal = {Nat. Nanotechnol.},
    author = {Froning, Florian N. M. and Camenzind, Leon C. and van der Molen, Orson A. H. and Li, Ang and Bakkers, Erik P. A. M. and Zumb{\"{u}}hl, Dominik M. and Braakman, Floris R.},
    number = {3},
    month = {3},
    pages = {308--312},
    volume = {16},
    doi = {10.1038/s41565-020-00828-6},
    issn = {1748-3387}
}

@article{Philips2022UniversalSilicon,
    title = {{Universal control of a six-qubit quantum processor in silicon}},
    year = {2022},
    journal = {Nature},
    author = {Philips, Stephan G. J. and M{\c{a}}dzik, Mateusz T. and Amitonov, Sergey V. and de Snoo, Sander L. and Russ, Maximilian and Kalhor, Nima and Volk, Christian and Lawrie, William I. L. and Brousse, Delphine and Tryputen, Larysa and Wuetz, Brian Paquelet and Sammak, Amir and Veldhorst, Menno and Scappucci, Giordano and Vandersypen, Lieven M. K.},
    number = {7929},
    month = {9},
    pages = {919--924},
    volume = {609},
    publisher = {Nature Research},
    url = {https://www.nature.com/articles/s41586-022-05117-x},
    doi = {10.1038/s41586-022-05117-x},
    issn = {0028-0836}
}

@article{schirk2025subharmonicfluxonium,
  title = {Subharmonic Control of a Fluxonium Qubit via a {Purcell-protected} Flux Line},
  author = {Schirk, J. and Wallner, F. and Huang, L. and Tsitsilin, I. and Bruckmoser, N. and Koch, L. and Bunch, D. and Glaser, N. J. and Huber, G. B. P. and Knudsen, M. and Krylov, G. and Marx, A. and Pfeiffer, F. and Richard, L. and Roy, F. A. and Romeiro, J. H. and Singh, M. and S\"odergren, L. and Dionis, E. and Sugny, D. and Werninghaus, M. and Liegener, K. and Schneider, C. M. F. and Filipp, S.},
  journal = {PRX Quantum},
  volume = {6},
  issue = {3},
  pages = {030315},
  numpages = {19},
  year = {2025},
  month = {Jul},
  publisher = {American Physical Society},
  doi = {10.1103/yx15-jyl7},
  url = {https://link.aps.org/doi/10.1103/yx15-jyl7}
}

@article{wang2024EDSRSihole,
  title = {Electrical operation of hole spin qubits in planar {MOS} silicon quantum dots},
  author = {Wang, Zhanning and Sarkar, Abhikbrata and Liles, S. D. and Saraiva, Andre and Dzurak, A. S. and Hamilton, A. R. and Culcer, Dimitrie},
  journal = {Phys. Rev. B},
  volume = {109},
  issue = {7},
  pages = {075427},
  numpages = {18},
  year = {2024},
  month = {Feb},
  publisher = {American Physical Society},
  doi = {10.1103/PhysRevB.109.075427},
  url = {https://link.aps.org/doi/10.1103/PhysRevB.109.075427}
}

@misc{simphony,
  author       = {{Qutility @ Faulhorn Labs}},
  title        = {Simphony: An open-source {Python} package for simulating spin dynamics in central spin systems},
  year         = {2026},
  publisher    = {GitHub},
  version      = {0.2.0},
  url          = {https://github.com/faulhornlabs/simphony},
  note         = {Accessed: 2026-07-13}
}

@article{rabi1937rabioscillation,
  title = {Space Quantization in a Gyrating Magnetic Field},
  author = {Rabi, I. I.},
  journal = {Phys. Rev.},
  volume = {51},
  issue = {8},
  pages = {652--654},
  numpages = {0},
  year = {1937},
  month = {Apr},
  publisher = {American Physical Society},
  doi = {10.1103/PhysRev.51.652},
  url = {https://link.aps.org/doi/10.1103/PhysRev.51.652}
}

\end{document}